\documentclass[10pt,english,journal]{IEEEtran}
\usepackage[english]{babel}
\usepackage[T1]{fontenc}
\usepackage[latin9]{inputenc}
\usepackage{geometry}
\usepackage{amsmath, amssymb, amsthm}
\usepackage{graphicx}
\usepackage[ruled,vlined,linesnumbered]{algorithm2e}
\usepackage{hyperref}
\usepackage{siunitx}
\usepackage{booktabs}
\usepackage{array}
\usepackage{caption}
\usepackage{xpatch}

\usepackage{esint}
\usepackage{mathtools}

\usepackage{nicefrac}
\usepackage{bigints}
\usepackage{mathtools}
\usepackage{blkarray, bigstrut}
\usepackage{calligra}

\usepackage{epstopdf}
\usepackage{dsfont}
\usepackage{ragged2e}
\usepackage{enumitem}
\usepackage{etoolbox}
\usepackage{babel}
\usepackage[nopar]{lipsum}
\usepackage{psfrag}
\usepackage{balance}
\usepackage{float}

\SetKw{KwBy}{by}

\makeatletter
\newcommand{\removelatexerror}{\let\@latex@error\@gobble}
\makeatother

\graphicspath{ {Figures/} }

\xpatchcmd{\proof}{\hskip\labelsep}{\hskip5\labelsep}{}{}  
\makeatletter
\xpatchcmd{\proof}{\@addpunct{.}}{\@addpunct{:}}{}{}
\makeatother

\renewcommand\[{\begin{equation}}
\renewcommand\]{\end{equation}} 
\usepackage{listings}
\usepackage{fancyvrb}
\usepackage{framed}

\usepackage{courier}
\usepackage[usenames,dvipsnames,table]{xcolor}

\definecolor{dkgreen}{rgb}{0,0.3,0}
\definecolor{gray}{rgb}{0.5,0.5,0.5}

\makeatletter
\newcommand*{\rom}[1]{\expandafter\@slowromancap\romannumeral #1@}
\makeatother

\usepackage{tabu}

\usepackage{multirow}
\usepackage{capt-of}

\usepackage{arydshln}
\newcommand{\comment}[1]{}
\usepackage{amsmath, amssymb, algpseudocode}

\usepackage{tabularx}
\usepackage{colortbl}
\usepackage{bbm}
\usepackage{mathrsfs}
\usepackage{cite}

\makeatletter
\patchcmd{\@maketitle}
  {\addvspace{0.5\baselineskip}\egroup}
  {\addvspace{-1\baselineskip}\egroup}
  {}
  {}
\makeatother

\definecolor{codegreen}{rgb}{0,0.6,0}
\definecolor{codegray}{rgb}{0.5,0.5,0.5}
\definecolor{codepurple}{rgb}{0.58,0,0.82}
\definecolor{backcolour}{rgb}{0.95,0.95,0.92}

\lstdefinestyle{mystyle}{
    backgroundcolor=\color{backcolour},   
    commentstyle=\color{codegreen},
    keywordstyle=\color{magenta},
    numberstyle=\tiny\color{codegray},
    stringstyle=\color{codepurple},
    basicstyle=\ttfamily\footnotesize,
    breakatwhitespace=false,         
    breaklines=true,                 
    captionpos=b,                    
    keepspaces=true,                 
    numbers=left,                    
    numbersep=5pt,                  
    showspaces=false,                
    showstringspaces=false,
    showtabs=false,                  
    tabsize=2,
    morekeywords={./mgen, instance, event, ON, UDP, DST, BURST, RANDOM, POISSON, EXP},
}

\begin{document}
\title{
Toward Explainable Reasoning in 6G: A Proof of Concept Study on Radio Resource Allocation
}
\author{
Farhad~Rezazadeh,~\IEEEmembership{Member,~IEEE}, Sergio Barrachina-Mu\~noz,~Hatim~Chergui,~\IEEEmembership{Senior~Member,~IEEE},\\~Josep~Mangues, Mehdi~Bennis,~\IEEEmembership{Fellow,~IEEE},~Dusit~Niyato,~\IEEEmembership{Fellow,~IEEE},~Houbing~Song,~\IEEEmembership{Fellow,~IEEE},\\~and~Lingjia~Liu,~\IEEEmembership{Senior~Member,~IEEE}

\IEEEcompsocitemizethanks{\IEEEcompsocthanksitem F. Rezazadeh and S. Barrachina-Mu\~noz are with the Telecommunications Technological Center of Catalonia (CTTC), Barcelona, Spain (e-mail: frezazadeh@cttc.es, sbarrachina@cttc.es).}
\IEEEcompsocitemizethanks{\IEEEcompsocthanksitem Hatim Chergui is with the i2CAT Foundation, Barcelona, Spain (e-mail: chergui@ieee.org)}

\IEEEcompsocitemizethanks{\IEEEcompsocthanksitem Josep Mangues is with the Telecommunications Technological Center of Catalonia (CTTC), Spain (e-mail: jmangues@cttc.es)}

\IEEEcompsocitemizethanks{\IEEEcompsocthanksitem Mehdi Bennis is with the University of Oulu, Oulu, Finland (e-mail: mehdi.bennis@oulu.fi)}

\IEEEcompsocitemizethanks{\IEEEcompsocthanksitem D. Niyato is with the School of Computer Science and Engineering, Nanyang Technological University, Singapore (e-mail: dniyato@ntu.edu.sg)}

\IEEEcompsocitemizethanks{\IEEEcompsocthanksitem H. Song is with Department of Information Systems, University of Maryland, Baltimore County (UMBC), Baltimore, USA (e-mail: h.song@ieee.org).}

\IEEEcompsocitemizethanks{\IEEEcompsocthanksitem L. Liu is with Bradley Department of Electrical and Computer Engineering, Virginia Tech, Blacksburg, USA (e-mail: ljliu@vt.edu).}
}

\maketitle
\thispagestyle{empty}

\begin{abstract}
The move toward artificial intelligence (AI)-native sixth-generation (6G) networks has put more emphasis on the importance of explainability and trustworthiness in network management operations, especially for mission-critical use-cases. Such desired trust transcends traditional post-hoc explainable AI (XAI) methods to using contextual explanations for guiding the learning process in an \emph{in-hoc} way. This paper proposes a novel graph reinforcement learning (GRL) framework named \emph{TANGO} which relies on a \emph{symbolic subsystem}. It consists of a \emph{Bayesian-graph neural network (GNN) Explainer}, whose outputs, in terms of edge/node importance and uncertainty, are periodically translated to a logical GRL reward function. This adjustment is accomplished through defined symbolic reasoning rules within a \emph{Reasoner}. Considering a real-world testbed proof-of-concept (PoC), a gNodeB (gNB) radio resource allocation problem is formulated, which aims to minimize under- and over-provisioning of physical resource blocks (PRBs) while penalizing decisions emanating from the uncertain and less important edge-nodes relations. Our findings reveal that the proposed \emph{in-hoc explainability} solution significantly expedites convergence compared to standard GRL baseline and other benchmarks in the deep reinforcement learning (DRL) domain. The experiment evaluates performance in AI, complexity, energy consumption, robustness, network, scalability, and explainability metrics. Specifically, the results show that TANGO achieves a noteworthy accuracy of 96.39\%  in terms of optimal PRB allocation in inference phase, outperforming the baseline by 1.22$\times$.

\end{abstract}


\maketitle

\section{INTRODUCTION}
\IEEEPARstart{6}{G} wireless technology is envisioned to revolutionize communication networks with ultra-reliable low latency communications (URLLC), enhanced mobile broadband (eMBB), and massive machine-type communications (mMTC) services. 
In this context, allocating radio resources in 6G networks for applications with diverse requirements is challenging. It requires adaptive policies to balance the distribution of resources effectively and meet corresponding performance criteria such as high data rates and low latency. Owing to the non-stationary nature of 6G networks, non-learning approaches struggle to provide accurate solutions under massive and complex 6G environments, where prediction modeling encounters numerous constraints~\cite{6GLetaief, 6GYang, STEP}. In contrast, deep reinforcement learning (DRL) emerges as a potential solution to be integrated within the 5G/6G networks~\cite{DRLQin, explWu, explLee, explZhou, explDai}. 
In addition to optimization methods based on DRL, some research suggests using graph models~\cite{GNNGu, GNNZhang, GNNJianzhe, GNNAsheralieva, GNNMengyuan, GNNNaderiAlizadeh, Chen_resource} to approach optimization problems by leveraging graph embedding techniques~\cite{GNNWang, GNNZhangRui, GNNCai}. Unlike conventional neural networks, GNNs operate by capturing relationships within graphs through message passing~\cite{GNNVignac} between their constituent nodes.
~Furthermore, in accordance with the European Commission's technical report on "Ethics Guidelines for Trustworthy AI"~\cite{XAI-Ethref}, the proposed AI solutions are encouraged to embody trustworthiness. Although AI/machine learning (ML) approaches have shown promising results and performance, concerns persist regarding the fundamental nature of deep neural networks (DNNs), often perceived as opaque models~\cite{Mosaheb_ref}.
In this regard, XAI is considered a promising approach to surmount the opaqueness of black-box models, enhancing transparency in the predictions and decision-making processes. Indeed, this explainability identifies the factors that positively or negatively affect the model's output. From our perspective, XAI approaches can be categorized into ante-hoc~\cite{xairefKok}, in-hoc~\cite{xairefSwamy, xairefRoy}, and post-hoc~\cite{xairefKalakoti, xairefAli}. 

Within the scope of our research, we extend the boundaries of XAI by introducing a novel approach to \emph{in-hoc} explainability~\cite{explrefSun, SliceOps}~(See Section~\ref{Subsec:In-hoc-xai}) solutions based on a symbolic subsystem. Here, XAI techniques can be utilized to incorporate explanations through regularization, supervision, or intervention mechanisms~\cite{exprefGao} directly into the training phase of GRL. The proposed symbolic subsystem benefits from a \emph{learning-reasoning}~\cite{refYunesyai} approach~(See Section~\ref{subsec:L-R}). The method innovatively integrates \emph{learning-for-reasoning}~\cite{refDinunesyai}~(See Section~\ref{subsec:L-F-R}) and \emph{reasoning-for-learning}~\cite{refRenkhoffnesyai}~(See Section~\ref{subsec:R-F-L}) concepts, creating a symbiotic relationship to reinforce each other. 
Furthermore, we aim to set a new benchmark in 5G network research and testing for \emph{radio resource allocation problems}, offering a comprehensive platform for exploring and validating next-generation network concepts and technologies.

\subsection{Related Works}
\label{sec:relatedwork}
Recent works such as~\cite{GRLRefShao, GRLOrhan, GRLREFSimsek, GRLREFDong, GRLREFDu, GRLREFXie} present different AI/ML approaches, including techniques based on GRL, to tackle different resource allocation sub-problems. 
Authors in~\cite{GRLRefShao} proposed a graph attention network (GAT) that enhances cooperation among base stations (BSs) in dense cellular networks. This aims to identify varying service demands both temporally and spatially, integrating this with mainstream DRL algorithms to develop an intelligent resource management strategy for network slicing. In~\cite{GRLOrhan}, the authors unveiled a connection management strategy for open RAN (O-RAN) radio intelligent controller (RIC) architecture utilizing GNN and DRL. The approach leverages the graph structure of O-RAN for neural network architecture, using RL for parameter learning. It focuses on balancing network traffic load and maximizing throughput. The work in~\cite{GRLREFSimsek} proposed a novel method using DRL and graph embedding to optimize network topology and maximize network capacity in dense cellular networks. In~\cite{GRLREFDong}, the authors introduced a novel network slicing and routing mechanism. They proposed a graph convolutional network (GCN) enhanced multi-task DRL solution to address the complexities of resource allocation effectively. The authors in~\cite{GRLREFDu} investigated the problem of resource management in mobile subnetworks and proposed a distributed multi-agent DRL model to reduce transmission failure rates. Additionally, it utilizes an attention-based GNN to identify potential interference among subnetworks to learn better policies while reducing computational complexity. The authors in~\cite{GRLREFXie} developed a novel neural network structure GNN with DRL, accompanied by an efficient algorithm for placing virtualized network function forwarding graphs (VNF-FGs). This approach optimizes resource allocation for virtualized network functions (VNF). 

\subsection{Gaps Analysis and Key Contributions}
\label{sec:technical_contribution}

\textbf{Lack of Transparency in Existing Approaches:} Previous studies~\cite{GRLRefShao, GRLOrhan, GRLREFSimsek, GRLREFDong, GRLREFDu, GRLREFXie} did not address the challenges of transparent and trustworthy resource allocation in 5G/6G networks. This gap in research is particularly critical given the significant concerns surrounding the inherent opacity and lack of transparency in DNN architectures. Such opaque models impede the establishment of trust in the decisions of the resource allocation agents, which becomes problematic when high reliability and trust are crucial to operating high-reliable 6G services.\\
\textbf{Limitations of Post-hoc XAI Techniques:} Current XAI techniques use methods like saliency maps~\cite{refMundhenknesyai} and feature attribution~\cite{refWangxai} explanations to explain neural network decisions.
Instead of these \emph{post-hoc} approaches, exploiting explanations during learning via symbolic rules can help impose domain constraints for trustworthiness and provide explicit reasoning traces for explainability~\cite{NeSySheth}. The TANGO framework uses a heuristic-based reward-shaping approach to incorporate domain-specific knowledge into the reward function through predefined rules, improving the learning process in three primary ways. First, it integrates domain constraints to ensure that the agent's actions adhere to overarching rules and long-term safety, even if they do not immediately optimize performance metrics. Second, the reward is shaped to guide the agent away from local optima and toward globally optimal strategies by reflecting long-term goals. Finally, symbolic rules enhance learning stability and speed by providing context that helps the agent prioritize effective and explainable actions, thereby reducing the exploration of suboptimal strategies. This allows network reconfiguration actions with higher confidence in real-world scenarios where the heterogeneity of services' workloads coupled with the multi-tier ultra-dense radio access network makes PRBs allocation a more complex task.\\ 
\textbf{Innovative Integration of Graph-based Representations and Bayesian Modeling:} In this direction, we propose \emph{TANGO}, a framework that combines graph-based representations and Bayesian modeling to decipher complex relationships between network states, aiming to optimally allocate resources even in diverse service scenarios. TANGO can continuously learn from the dynamic network environment, adapting its PRB allocation strategies in real time to follow the changing traffic patterns. Our contributions can bridge the gap between graph theory and RL and lay the groundwork for developing more understandable and efficient GRL agents in complex 6G settings. This innovative integration is designed to adeptly address the challenges associated with structured or relational state representations. Unlike traditional RL policies that operate on flat vector-based state representations, \emph{TANGO} RL policy operates directly on graph-like data structures. This architecture inherently enables the model to discern the relational information present within network states. This paper presents the following contributions:

\begin{itemize}

\item It introduces a comprehensive 5G stand-alone testbed architecture (See Section~\ref{Sec:testbed_architecture}). Central to this architecture is an application programming interface (API) client developed for gNB interactions, a cloud-native platform for monitoring and gathering RAN metrics, an open-source traffic generator, and a decision engine incorporating AI/ML algorithms. 

\item We specifically introduce the \emph{TANGO} framework, a GRL solution integrated into the testbed's decision engine. \emph{TANGO} transforms the network's state space into a scalable graph format, targeting efficient PRBs allocation.

\item TANGO augments GNN-based REINFORCE~\cite{REINFORCEZhang} (See Section~\ref{Sec:REINFORCEbaseline}) with a return baseline to mitigate high variance~\cite{REINFORCEChen} and introduces advantage normalization to stabilize the training. Furthermore, it incorporates dropout-based~\cite{REINFORCEGuo} regularization and layer normalization~\cite{REINFORCECai} into the GNN architecture as well as a learning rate scheduler~\cite{REINFORCEXiong} ~(See Section~\ref{Sec:GNN-REINFORCEtech}). Experimentally, we have observed that fine-tuning these techniques speeds up the convergence process compared to benchmarks (See Section~\ref{sec:ai_met}), making the approach more stable and practical for real-world applications.

\item Given that GNN-REINFORCE algorithm lacks transparent reasoning ~\cite{refShengzhenesyai} and explainability, TANGO includes a symbolic subsystem with \emph{Bayesian-GNN explainer} and \emph{reasoner} modules.

\item This \emph{Bayesian-GNN explainer} (See Section~\ref{Sec:TANGO_Explainer}) employs variational Bayesian inference~\cite{bayesrefFox, bayesrefWan, bayesrefZhang} to approximate the posterior distribution of the masks~\cite{explAmara, explFunke}, highlighting the importance of each edge and node feature in the graph using observed data. This allows introspection~\cite{ExplainabilityDuval, ExplainabilityVu, ExplainabilityHe, ExplainabilityLuo} into the GRL agent's decision-making process, which highlights how each network state contributes to predictions (e.g., the amount of required radio resources) and enables taking informed actions.

\item The \emph{Reasoner} (See Section ~\ref{Sec:TANGORewardShaper}) approach~\cite{r-shapingref, rewardshrefKopic} is designed to manage and execute predefined logical rules on perceived node and edge importance and associated uncertainty scores~\cite{uncerrefRuah, uncerrefZecchin}, obtained from the Bayesian explainer, whereby each rule comprises a condition and an action. It enables agents to precede decisions that impact crucial graph nodes or relationships. This approach provides additional reward feedback, particularly in 6G network environments characterized by sparse or deceptive rewards, thereby preventing entrapment in local optima.

\item  Finally, the paper provides an in-depth examination of performance~(See Section~\ref{sec:perf_eval}) across various metrics, encompassing AI efficiency, complexity, energy consumption, robustness, network performance, scalability, and explainability. The PoC study proves the superiority of TANGO by achieving an outstanding accuracy of 96.39\%, outperforming the baseline
with 1.22$\times$ radio resource allocation improvement. TANGO speeds up the convergence process compared to the baseline method and other benchmarks through a custom-designed OpenAI Gym toolkit~\cite{Brockman_gym, Globe_far}.
\end{itemize}

The remainder of this paper is organized as follows: Section~\ref{Sec:Preliminaries} elaborates preliminaries and introduces in-hoc XAI and neuro-symbolic AI for 6G networks. Section~\ref{Sec:testbed_architecture} delves into the 5G testbed architecture and components. Section~\ref{sec:Optimization} formulates the radio resource allocation problem and describes the scenario under consideration. Then, Section~\ref{sec:TANGO_Framework} highlights how various components of the \emph{TANGO} frameworks interplay. The article validates the proposed approach's efficacy and performance compared to the standard GRL baseline and other benchmarks in Section~\ref{sec:perf_eval}. Finally, Section~\ref{sec:concusion_future} provides the concluding remarks and outlines the future work.

\section{Preliminaries}
\label{Sec:Preliminaries}
\subsection{In-hoc Explainable Artificial Intelligence in 6G Networks}
\label{Subsec:In-hoc-xai}
The evolution toward 6G networks emphasizes the need for reliable AI. In this respect, the importance of XAI has been emphasized in ensuring critical services, enhancing security, and addressing resource allocation \cite{root, trustDL}. 
Explanation-guided learning (EGL) emerges as a promising \emph{in-hoc} paradigm, aiming to enhance model trustworthiness and performance by incorporating XAI-driven signals during training \cite{egl}. This approach offers a principled way to preserve model performance while ensuring explainability and reducing bias, thanks to its typical problem formulation,
\begin{subequations}
\label{OPT1}
\begin{equation}
\min\,\underbrace{\mathcal{L}_{Pred}\left(\mathcal{F}\left(\mathcal{X}\right), \mathcal{Y}\right)}_{prediction}+ \underbrace{ \alpha\mathcal{L}_{Exp}\left({g}\left(\mathcal{F},\left(\mathcal{X,Y}\right)\right), \mathcal{M}\right) }_{explanation} 
\end{equation}

\begin{equation}
     \mathrm{s.t.}\hspace{5mm}\rho_{k,n} \geq \gamma_{n}\label{recall},
\end{equation}
\end{subequations}
consisting of \textit{i)} the main loss term related to the prediction/action objective $\mathcal{L}_{Pred}$ of model $\mathcal{F}$, where $\mathcal{F}$ denotes the model or function that maps the input features$\mathcal{X}$ to predictions $\mathcal{Y}$. \textit{ii)} the loss supervising the model explanation $\mathcal{L}_{Exp}$ using masked dataset $\mathcal{M}$, where $\mathcal{M}$ represents a masked version of the dataset, which is used in the explanation loss $\mathcal{L}_{Exp}$ and  $\mathcal{X,Y}$ stand for the input features and predictions, respectively. The explanation loss makes use of the XAI raw attributions (e.g., SHAP and Integrated Gradients) or composite metrics such as \emph{confidence score, log-Odds, comprehensiveness, entropy} \cite{brik2023survey, tradeoff, egdrl} to control the trustworthiness of the AI model. These metrics, along with the predictions/actions, are iteratively exchanged between the \emph{explainer} and the model's \emph{optimizer} in a closed-loop fashion to guide the learning in run-time. In practice, the XAI feedback signal has various forms. It can measure the model trustworthiness via a KL divergence distance between original predictions and those emanating from an XAI-based masked dataset. Moreover,
It can also be an extra reward signal in DRL \cite{SliceOps} or a logical reasoning term in neuro-symbolic AI formulations, as explained in the following. Finally, \textit{iii)} a recall constraint might be considered to enforce the desired fairness level \cite{fair}.

\subsection{Neuro-Symbolic AI in 6G Networks }
Neuro-symbolic AI (NeSy AI) is a technology that combines neural networks and symbolic methods to enhance AI capabilities. It aims to replicate human cognitive processes by understanding and manipulating symbols. This integration leads to better performance in areas such as solving complex wireless problems~\cite{refShengzhenesyai, refShirajumnesyai,refThomasumnesyai}. Utilizing symbols allows for a transparent and understandable way to approach logical reasoning. However, these systems can often lack flexibility and a broad range of applicability. By combining these symbolic approaches with neural networks, we can take advantage of the advances in deep learning, which promotes more effective knowledge acquisition and improved generalization~\cite{refBajajnesyai}. XAI does not necessarily aim to subject AI to human standards. DNNs follow a bottom-up approach, where the network draws inferences and generalizations from data points. On the other hand, the NeSy AI approach is a top-down method that starts with reasoning. This approach can potentially bridge the gap between explainability and performance by building inherently more interpretable models. Various studies have explored the current applications of NeSy AI, including the studies in~\cite{refRenkhoffnesyai, refAcharyanesyai, refDinunesyai, refYunesyai}. These applications are categorized into three main models, as discussed below.

\subsubsection{Learning for Reasoning}
\label{subsec:L-F-R}
The objective is to leverage the strengths of symbolic systems and neural networks, enhancing the efficiency and effectiveness of problem-solving processes. DNNs simplify the search space in symbolic systems, improving computational efficiency. Additionally, reasoning is integrated into the learning process as a regularization method. Symbolic knowledge provides a framework that guides machine-learning tasks. This dual approach involves applying neural networks to refine symbolic reasoning and extracting symbolic representations from data using neural networks. Essentially, neural networks act as tools for acquiring knowledge that aids in symbolic reasoning tasks. The process starts with a neural network layer, $\mathcal{F}_{\theta}(\mathcal{X})$, which learns from input $\mathcal{X}$ to produce output $\mathcal{Y}$. Here, $\theta$  represents the parameters of the neural network. This layer adjusts its parameters through learning methods such as backpropagation. Then, a symbolic reasoning layer, $R_{rule}$, is applied, which ensures that the output $\mathcal{Y}$ and the features of $\mathcal{X}$ comply with specific logical rules. The expression representing this process is given as follows,
\begin{equation}
    \mathcal{Y} = \mathcal{F}_{\theta}(\mathcal{X}) \quad \text{and} \quad R_{rule}(\mathcal{Y}, \text{features of } \mathcal{X}) = \text{True}.
\end{equation}

This dual-layer approach makes the model suitable for tasks requiring both data analysis and strict logical adherence.

\subsubsection{Reasoning for Learning}
\label{subsec:R-F-L}
In this model, there is a shift in the traditional roles played by symbolic and sub-symbolic elements. The sub-symbolic component plays a more prominent role in problem-solving, while the symbolic part supports the neural network. Using symbolic knowledge can significantly enhance high-level reasoning and decision-making in complex systems like 6G networks, where the environment is often unpredictable and constantly changing. The rationale behind reasoning for learning is to harness the power of neural systems for machine learning tasks while incorporating symbolic knowledge to improve performance and interpretability. Symbolic knowledge is typically encoded in a format suitable for neural networks and used to guide or constrain the learning process. An example of this is encoding symbolic knowledge as a regularization term in the loss function of a specific task. This approach utilizes symbolic constraints, referred to as $\mathcal{C}$, to direct and shape the learning objective. The learning objective incorporates the standard loss function with an additional component that takes into account compliance with these symbolic constraints,
\begin{equation}
    \mathcal{L}(\theta) = \text{Original-Loss}(\mathcal{F}_{\theta}(\mathcal{X}), \mathcal{Y}) + \lambda \cdot R_{rule}(\mathcal{C}, \mathcal{F}_{\theta}(\mathcal{X})).
    \label{eq:r-f-l}
\end{equation}
In Equation~\ref{eq:r-f-l}, $\text{Original-Loss}(\mathcal{F}_{\theta}(\mathcal{X}), \mathcal{Y})$ represents the typical loss function used in neural networks, which aims to minimize the discrepancy between the network's predictions and the actual outputs. The second term, $\lambda \cdot R_{rule}(\mathcal{C}, \mathcal{F}_{\theta}(\mathcal{X}))$, introduces symbolic reasoning, where $R_{rule}$ evaluates the network's predictions against the symbolic constraints $\mathcal{C}$. The parameter $\lambda$ is crucial in determining the balancing between the original loss and the symbolic constraints.

\subsubsection{Learning-Reasoning}
\label{subsec:L-R}
The learning-reasoning category represents a powerful combination of learning and reasoning, where both work together to enhance problem-solving capabilities. This approach involves a dynamic and reciprocal interaction between symbolic and sub-symbolic elements, where each component's output informs the input of the other. The main objective of learning-reasoning is to strike a harmonious balance between the involvement of neural and symbolic systems in the problem-solving process. The neural network's output is utilized as input to the symbolic reasoning component, and the output of symbolic reasoning is used as input to the neural network. The joint objective function presented in this approach is a combination of two interconnected components,
\begin{equation}
    \mathcal{J}(\theta, R_{rule}) = \mathcal{L}(\theta, \mathcal{X}, \mathcal{Y}) + \gamma \cdot \mathcal{S}_{sy}(R_{rule}, \mathcal{F}_{\theta}(\mathcal{X})).
\end{equation}
The first term, $\mathcal{L}(\theta, \mathcal{X}, \mathcal{Y})$, is responsible for minimizing prediction errors during the neural network's conventional learning process. The second term, $\mathcal{S}_{sy}(R_{rule}, \mathcal{F}_{\theta}(\mathcal{X}))$, evaluates the effectiveness of reasoning by integrating symbolic reasoning and ensuring that the neural network's predictions comply with logical rules denoted by $R_{rule}$. The value of the parameter $\gamma$ plays a significant role in determining the importance of learning and reasoning in the model's objective. This integration helps to balance data-driven learning and logical consistency, thereby leveraging the strengths of both neural networks and symbolic logic. 

\vspace{-.2cm}
\subsection{Symbolic Reward shaping}
The RL reward function plays a crucial role in clearly defining the agent's objectives. It transforms the task's goals into a numerical form that the agent can understand and execute. The effectiveness of the agent's actions is evaluated based on their alignment with the objectives, which is determined by the rewards or penalties received from the designed reward function~\cite{symrewZhou}. In this context, reward shaping~\cite{rewshpingSaha, SliceOps,egdrl} is an effective technique that enhances RL by providing additional feedback to the agent. Symbolic reward shaping~\cite{refAcharyanesyai, symrewBougie} can be a hybrid \emph{learning-reasoning} approach that leverages the strengths of both ML and symbolic AI. It identifies significant symbolic events or rules that trigger supplementary rewards. These intermediate goals act as a guide to help the agent achieve the task objective, especially in environments with sparse or delayed rewards like 6G~\cite{sparsreLi, sparsreHuang, sparsreSun, sparsreWenhan}. By combining traditional RL with symbolic AI, reward shaping integrates domain-specific knowledge and desired behaviors into the reward signal through symbolic logic. It can lead to more robust, explainable, and effective outcomes in complex environments where learning from raw data and reasoning through predefined logic are crucial. Indeed, the system learns optimal behaviors through interaction while simultaneously applying logical reasoning to guide these interactions. However, the design of the potential function and corresponding symbolic rewards is critical. Poorly designed reward structures can negatively affect the RL agent's learning. 

Symbolic reward shaping can employ symbolic rules with a logical structure to make decision-making processes interpretable, explicit, and directly influenced by human expertise. These rules are explicit \emph{if-then} statements that encode logic in a human-readable and modifiable form. Symbolic rules are beneficial in RL~\cite{refAcharyanesyai} to embed domain-specific knowledge into the learning process, directing the agent towards beneficial behaviors that may not be immediately obvious or rewarding based on the environment's natural dynamics alone. Symbolic rules can significantly impact GRL due to the complex relationships and attributes of nodes and edges in graphs. By modifying the reward signal based on additional criteria derived from domain knowledge or desirable behaviors, GRL can be optimized for improved outcomes. It can be implemented to manipulate or adjust rewards based on the attributes of nodes and edges and their topological configurations.

\section{Testbed Architecture}
\label{Sec:testbed_architecture}
\begin{figure}[t!]
\centering
\includegraphics[width=.8\columnwidth]{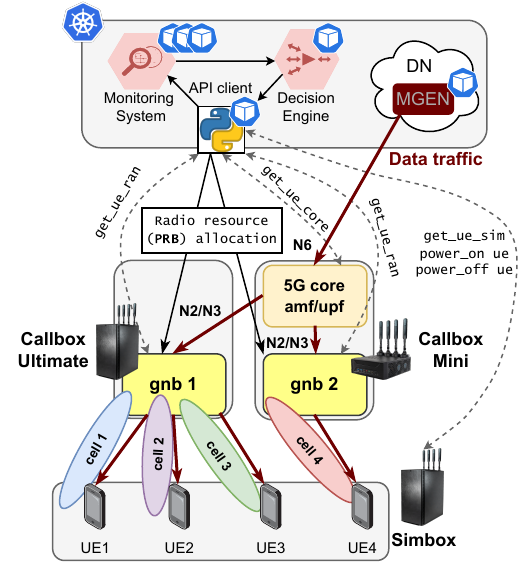}   
\caption{Infrastructure and network setup~\cite{rezazadeh2023x}.}
\label{fig:testbed_architecture}
\vspace{-.7cm}
\end{figure}

Figure~\ref{fig:testbed_architecture} illustrates the 5G stand-alone (SA) testbed considered in this work. It consists of specific hardware and software components to realize both the 5G infrastructure, i.e., RAN and core, and management-related functionalities, e.g., monitoring system and decision engine. For implementing the gNBs and the 5G core, the testbed relies on a Callbox Ultimate and a Callbox Mini by Amarisoft.
~To emulate user equipment (UE) interactions with the 5G network, the testbed utilizes a Simbox~\cite{AM_simbox}, also from Amarisoft. Equipped with multiple UE profiles and scenarios, Simbox can emulate diverse network conditions, from densely populated urban settings to remote areas with sporadic connectivity. The choice to integrate Amarisoft components was motivated by their reputation as an accurate research and testing solution. In our study, we employ a hardware-based testbed rather than relying exclusively on simulations. This approach ensured greater accuracy and practical relevance in our experimental findings, enabling us to capture the nuanced behaviors and challenges of real-world 5G/6G environments.

\begin{figure}[t!]
	\centering
\includegraphics[width=0.4\textwidth] {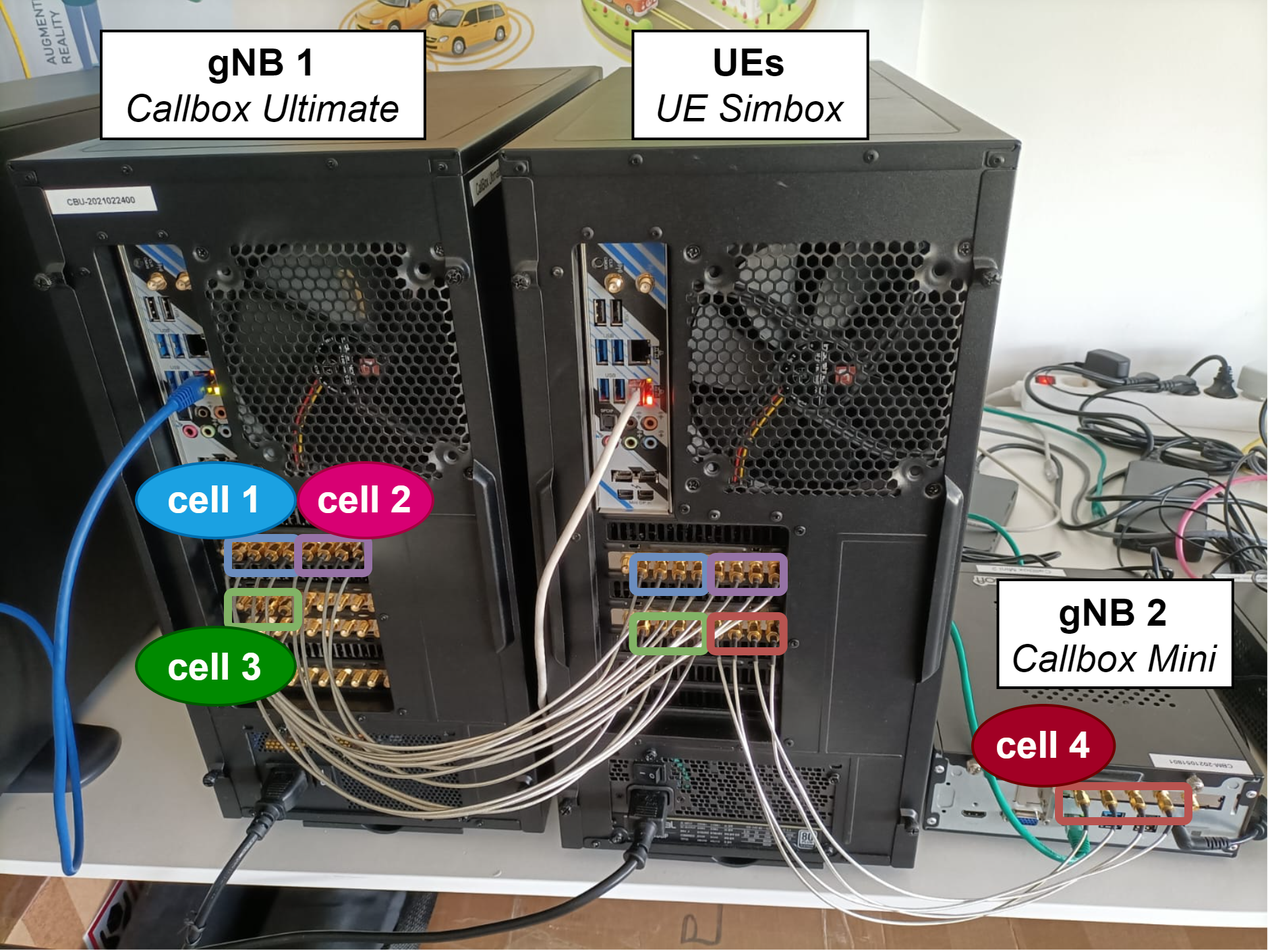}
	\caption{Amarisoft 5G RAN equipment.}\label{fig:amarisoft}
\vspace{-.7cm}
\end{figure}

While our current 5G testbed is predicated on the utilization of a gNB devoid of O-RAN capabilities, it is essential to emphasize the generalizability and scalability of our findings. The experiments conducted in this study are not intrinsically bound to the specific gNB architecture deployed. Instead, they embody broader principles applicable to the evolving landscape of O-RAN implementations. Figure~\ref{fig:amarisoft} shows the physical setup of the implemented 5G RAN. A wired radio frequency (RF) setup avoids undesired channel instabilities and ensures reproducibility.

In addition to the foundational 5G infrastructure, our testbed integrates a cloud-native platform for monitoring and management. This platform, rooted in flexibility, scalability, and resilience principles, leverages the power of Kubernetes, the de facto open-source container orchestration system. The platform has four main components: the API client, the monitoring system, the decision engine, and the traffic generator.

\textbf{API Client}: The RAN API client provides an intermediary interface with the gNBs.~
It fulfills two primary functions: first, it retrieves metrics from the RAN to supply the monitoring system, and second, it executes PRB-allocation actions as prompted by the decision engine. The API client is a custom Python solution that leverages the Callbox APIs hosted by the gNBs. These APIs facilitate the acquisition of real-time data spanning a plethora of KPIs (e.g., number of connected devices, bandwidth utilization, or channel quality) and enabling remote configuration (e.g., turning cell ON/OFF, UE handover, or PRB allocation). For the latter, the API call includes the IP address and port number of the target gNB, the target cell identifier, an allocation flag defining whether the resource block allocation is fixed, and the boundaries of the desired radio resource block allocation.

\textbf{Monitoring System}: The monitoring system systematically gathers and preprocesses data concerning the status of network service-supporting elements, which are often distributed across multiple domains. This system relies on sampling functions that request data from embedded element managers and publish it onto a Kafka bus (See~\cite{d33}). In this work, the monitoring system focuses on providing RAN metrics as a primary consideration, with future prospects for additional metrics encompassing cloud resource status (e.g., CPU utilization, cost metrics, and transport latency) to address higher dimensional optimizations. In summary, the monitoring system exposes the metrics for the decision engine's determination of whether to enact a configuration action.

\textbf{Decision Engine}: The decision engine extends beyond its conventional decision-making role by incorporating AI and ML algorithms at its core. It can continuously learn, adapt to evolving conditions, and fine-tune its operations for improved efficiency~\cite{Blanco_sc_ref}. One of its noteworthy capabilities is the ability to forecast network traffic demands accurately. By proactively adjusting PRB allocation in response to these predictions, the system enhances the network's overall bandwidth efficiency and performance.

\textbf{Traffic Generator}: Finally, a traffic generator is indispensable for testing the proposed framework. In this context, we rely on MGEN as a versatile utility to fulfill this need. MGEN facilitates IP network performance tests and measurements using TCP and UDP traffic. The toolset can generate a wide range of real-time traffic patterns so that the network can be loaded in various ways. This allows our testbed to undergo rigorous testing across various scenarios~(See Section~\ref{sec:perf_eval})~\cite{rezazadeh2023multi}.

\section{Use Case and Optimization Objective}
\label{sec:Optimization}
We cast the radio resource allocation problem as an optimization challenge, which aims to identify policies that minimize both the under-provisioning and over-provisioning of radio resources within the system's constraints. The Amarisoft system configures PRB allocation for each cell via various files, especially the RAN configuration file (e.g., enb.cfg
and gnb.cfg)\footnote{All testbed configuration files are open source. This encourages the sharing of code among contributors and allows researchers to improve upon the work of others under the GNU General Public License v3.0. \textit{The repositories will be disclosed upon acceptance}.}. This configuration specifies several parameters, including settings for the physical, medium access control (MAC), and radio resource control (RRC) layers, which influence how resources like PRBs are allocated. In such scenarios, let us consider minimizing under-provisioning and over-provisioning while maintaining quality of service (QoS) requirements. Table~\ref{tab:notation} presents a comprehensive summary of the parameters and variables characterizing the system. Let $f(a_t, s_t)$ represent the objective function, where $a_t$ is the action (PRB allocation) taken at time $t$ and $s_t$ is the network state at time $t$. We can define the objective function as follows:
\renewcommand{\arraystretch}{1}
\begin{table}[t!]
\caption{Major Mathematical Notations}
    \centering
    
    \begin{tabularx}{\columnwidth}{@{}l X@{}}
        \toprule
        \textbf{Symbol} & \textbf{Description} \\
        \midrule
        \( \tilde{G} \) & Graph representation of the network environment\\
        \( V \) & Set of nodes in the graph \( \tilde{G}\)\\
        \( E \) & Set of edges in the graph \( \tilde{G} \)\\
        \( \mathcal{A} \) & Action space of the RL agent\\
        \( \mathcal{A}_{\text{max}} \) & Maximum number of PRBs\\
        \( \mathcal{S} \) & State space of the network environment\\
        \( \mathcal{R} \) & Reward function\\
        \( \lambda_w \) & Weighting factor\\
        \( r \) & Reward received by the agent after taking an action in a state\\
        \(r^*\) & XAI reward\\
        \( \pi \) & Policy of the RL agent\\
        \(\Delta_{\text{over}}\) & Over-provisioning constraint\\
        \(\Delta_{\text{under}}\) & Under-provisioning constraint\\
        \( \theta \) & Parameters of the GNN-based policy\\
        \( \alpha \) & Learning rate for the optimization algorithm\\
        \( \gamma \) & Discount factor for the reinforcement learning algorithm\\
        \(T\) & Training duration\\
        \(T_e\) & Maximum timesteps per episode\\
        \(\tau \) & Periodicity of reward shaping\\
        \(G\) & Discounted return\\
        \(\bar{G}\) & Mean discounted return\\
        \(\sigma_G\) & Std deviation of discounted return\\
        \(\delta\) & Advantage with baseline\\
        \(H\) & Entropy bonus\\
        \(\lambda_H\) & Entropy weight \\
        \(\mathcal{L}\) & Total loss\\
        \(\mathcal{T}\) & Trajectory or experience per episode \\
        \(s_g\) & Graph representation of state $s$\\
        \(s_x\) & Node features of the graph\\
        \(s_i\) & Edge indices of the graph \\
        \(P_a\) & Probabilities of taking different actions $a$ given
        the state $s_g$\\
        \(\mathcal{I}^v_{\varrho}\) & Feature importance threshold\\
        \(\mathcal{I}^i_{\varrho}\) & Node importance threshold\\
        \(\mathcal{U}_{\varrho}\) & Uncertainty threshold \\
        \(\mathcal{Z}\) & Feature similarity (cosine similarity) 
        between connected nodes\\
        \(\mathcal{Z}_{\varrho}\) & Relationship strength threshold\\
        \(\phi\) & Log normalizing constant \\
        \(\mathbbm{1}\) & Indicator function\\
        \(\odot\) & Element-wise multiplication\\
        \bottomrule
    \end{tabularx}
    \label{tab:notation}
\end{table}
\begin{equation}
    f(a_t, s_t) = \text{Traffic}(s_t) - \lambda_w \cdot \text{Served-Traffic}(a_t, s_t),
\end{equation}
where $\emph{Traffic}(s_t)$ represents the amount of downlink traffic based on the current state $s_t$, and $\emph{Served-Traffic}(a_t, s_t)$ represents the actual allocated radio resources based on the action $a_t$ and the current state $s_t$. $\lambda_w$ is a weighting factor that balances traffic and served traffic. All values are fetched using the discussed API client in Section~\ref{Sec:testbed_architecture}. Here, the terms \emph{state}, \emph{action}, and \emph{reward} are defined within the framework of DRL~\cite{farha_drl_ref} and discussed in Section~\ref{Sec:REINFORCEbaseline}. The optimization problem can be formulated as finding the optimal action $a_t^*$ at each time step $t$ that minimizes the objective function $f(a_t, s_t)$. Mathematically, this can be expressed as,
\begin{equation}
    a_t^* = \arg \min_{a_t \in \mathcal{A}} \{ f(a_t, s_t) \}.
\end{equation}
Let~$\Delta_{\text{over}}$ constraint denote the maximum allowable over-provisioning margin, where the allocated resources must not exceed the required resources by a significant margin. Furthermore, $\Delta_{\text{under}}$ constraint represents the maximum allowable under-provisioning margin, ensuring that the allocated resources are adequate to meet the required traffic demand. The problem can be formalized as,

\noindent \textbf{Problem}~\texttt{Radio Resource Allocation}:
\begin{flalign}
\label{prob:RAN_Slicing}
\text{min} & \lim_{{T\rightarrow\infty}} \sum_{t = 0}^{T-1} \mathbb{E}\left[ f(a_t, s_t) \right] & & \\
\noindent\text{subject to:} & & & \nonumber \\
& f(a_t, s_t) \geq  \Delta_{\text{over}}, & \forall t \in T & \\
& f(a_t, s_t) \leq  \Delta_{\text{under}}, & \forall t \in T.
\end{flalign}
The optimization task can be addressed using the DRL framework. Within this DRL, the state and action spaces, as well as the reward function, are defined as follows,
\begin{figure*}[ht]
	\centering
	\includegraphics[width=.8\textwidth] {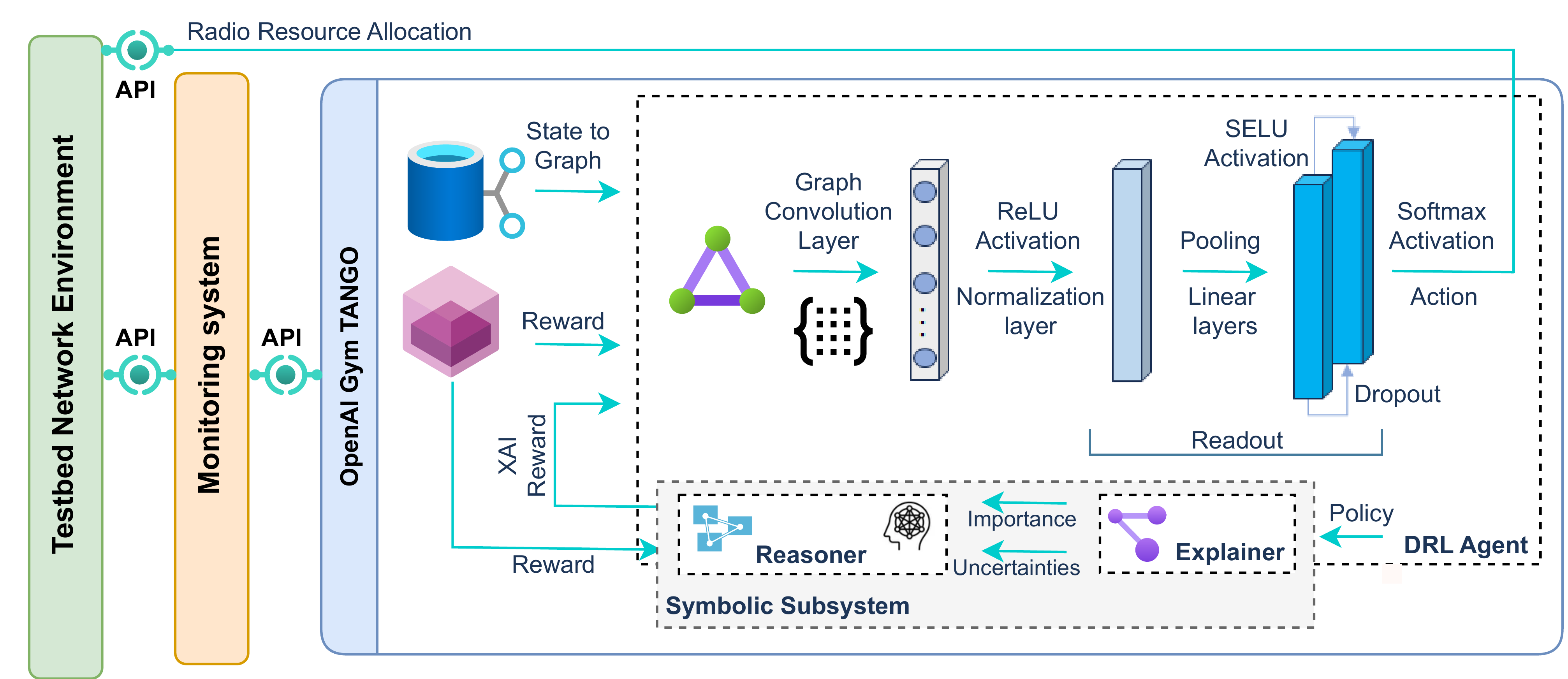}
	\caption{The TANGO framework workflow illustrates the process from network state transformation to decision-making and refining radio resource allocation policies.}\label{fig:TANGO-WORKFLOW}
 \label{fig:TANGO_workflow}
\end{figure*}

\textbf{Action Space $\mathcal{A}$}: Let  $\kappa$ denote the smallest PRB allocation increment, also referred to as the \emph{chunk size}. Consequently, our action space $\mathcal{A}$ can be represented as, 
\\
\begin{equation}
\mathcal{A} = \left\{ n\cdot \kappa : n \in \mathbb{Z}^+, 0 \leq n\cdot \kappa \leq \mathcal{A}_{\text{max}} \right\},
\end{equation}
where $\mathbb{Z}^+$ denotes the set of positive integers and $\mathcal{A}_{\text{max}}$ is the PRB capacity.

\textbf{State Space $\mathcal{S}$}: The state of the DRL agent is a four-dimensional representation at time $t$ that is constituted by,
\begin{enumerate}
    \item Signal-to-noise ratio (SNR): 
The SNR, represented as \( s_1^{t} \), quantifies the quality of a communication signal compared to the background noise. A superior SNR indicates enhanced signal clarity and improved transmission fidelity. 
   \item Traffic Volume: This metric represented by \( s_2^{t} \) indicates the amount of data transmitted or communicated over the network.
\item Residual Capacity: The leftover capacity ($s_3^{t}$) reflects the outstanding available capacity within the 5G network for radio resources. This metric is essential to manage additional data traffic or load without reaching saturation.
   \item Allocation Gap: It is denoted by $s_4^{t}$ and highlights the difference between the allocated and the required resources. Recognizing this gap is crucial for informed resource allocation decisions.
\end{enumerate}
Therefore, the state space of the DRL agent is given by:

\begin{equation}
\begin{aligned}
\mathcal{S}^{t} = [& s_1^{t} \; (\text{SNR}), \; s_2^{t} \; (\text{Traffic volume}), \\
& s_3^{t} \; (\text{Leftover capacity}), \; s_4^{t} \; (\text{Allocation gap}) ].
\end{aligned}
\end{equation}
The state space provides a complete insight into the network's behavior and prevailing conditions. These states enable the DRL agent to make informed decisions in the Amarisoft 5G testbed.

\textbf{Reward $\mathcal{R}$}:
The reward function $\mathcal{R}(s,a)$ is expressed as a piecewise function as in~\ref{Equ:rewrd_func}. The foundational component of the reward structure is denoted by $f(a_t, s_t)$, representing the base reward associated with choosing action $a_t$ in state $s_t$. This is complemented by bonus $\text{B}$ to encourage the GRL agent and penalties $P_{\text{over}}$ and $P_{\text{under}}$, which are imposed when the allocation of radio resources does not meet the predefined constraints.
~These elements form a reward function that balances achieving optimal resource utilization with adherence to critical operational constraints.

\begin{equation}
\mathcal{R}(s_t, a_t) = 
\begin{cases}
    f(a_t, s_t) + B, & \text{if } \left(f(a_t, s_t) \geq  \Delta_{\text{over}} \right. \\
             & \text{and}\left. f(a_t, s_t) \leq  \Delta_{\text{under}}\right)
    \\
    f(a_t, s_t) - P_{\text{over}}, & \text{if } \left( f(a_t, s_t) <  \Delta_{\text{over}} \right) \\
    f(a_t, s_t) - P_{\text{under}}, & \text{if } \left( f(a_t, s_t) >  \Delta_{\text{under}}\right)
\end{cases}
\label{Equ:rewrd_func}
\end{equation}


\section{TANGO Decision Engine}
\label{sec:TANGO_Framework}
In this section, we delve into the details of the \emph{TANGO} framework. To provide a visual perspective, we have illustrated the entire workflow in Figure~\ref{fig:TANGO_workflow}. This illustration depicts the interplay between various components and highlights the pivotal stages in the process. Moreover, to further simplify the comprehension of our approach, we have encapsulated the core essence of the \emph{TANGO} framework in Algorithm~\ref{algo:REINFORCE}.

\subsection{REINFORCE Algorithm}
\label{Sec:REINFORCEbaseline}
The REINFORCE algorithm is a DRL approach based on a policy gradient where an agent interacts with the testbed environment in the \emph{TANGO} framework over sequential time steps. At each step $t$, the agent observes a state $s_t$, decides on an action $a_t$ based on its policy $\pi$, receives a reward $r_t$, and then transitions to the next state $s_{t+1}$. In problem \ref{prob:RAN_Slicing}, the transition from the current state to the next state is directly influenced by the action taken by the agent, which is the allocation of PRBs. At each time step $t$, the state $s_t$ captures several network parameters, i.e., the SNR, traffic volume, residual capacity, and allocation gap. When the agent chooses an action $a_t$ and allocates a certain number of PRBs, the network is affected in the following ways. For example, allocating optimal PRBs helps to serve more traffic, reducing the allocation gap (the difference between the required and allocated resources) and potentially increasing the residual capacity if the demand is met efficiently. However, if the action results in over-provisioning (allocating more PRBs than needed) or under-provisioning (allocating fewer PRBs than required), it will lead to wasted resources or unserved traffic. This, in turn, updates the traffic load, residual capacity, and allocation gap in the next state $s_{t+1}$. Allocating PRBs more efficiently can reduce network congestion by better-distributing resources among users. The efficient utilization of resources adds degrees of freedom to the system and increases the admission rate. The agent aims to determine the optimal policy that maximizes the expected cumulative reward. This  cumulative reward is discounted by a factor $\gamma \in [0,1]$ and given by $G_t = \sum_{j=0}^{\infty} \gamma^k r_{t+j}$. Let us consider $\pi_\theta(a|s)$ as the probability that the agent takes action $a$ in state $s$, following the policy $\pi_\theta$.
The objective function can be written as, $J(\theta) = \mathbb{E}_{\pi_\theta} \left[ G_t | s_t = s \right]$. The REINFORCE method modifies the policy parameters following stochastic gradient ascent on expected return $J(\theta)$ where the gradient is given by,
\begin{equation}
    \nabla_\theta J(\theta) = \mathbb{E}_{\pi_\theta} \left[ G_t \cdot \nabla_\theta \log \pi_\theta(a_t|s_t) \right].
\end{equation}
\begin{algorithm}[t!]
    \caption{TANGO-GRL Algorithm}
    \label{algo:REINFORCE}
    \small
    \SetKwFunction{FMain}{REINFORCE}
    \SetKwFunction{FExplainer}{Explainer}
    \SetKwFunction{FReasoner}{Reasoner}
    \SetKwInput{KwRequire}{Require}
    
    \KwRequire{$\mathcal{S}$, $\mathcal{A}$, $\alpha$, $\gamma$, $\tau$, $\lambda_H$, $\mathcal{I}_{\varrho}$, $\mathcal{U}_{\varrho}$, $\mathcal{Z}_{\varrho}$~~(See Table~\ref{tab:notation})}
    \KwResult{Optimized Policy Parameters $\theta$}
    
    \tcp{Initialization Phase}
    Import testbed environment/interfaces (`TANGO--v1')\;
    Initialize policy $\pi(a|s;\theta)$ with parameters $\theta$\;
    Initialize symbolic reasoner and define symbolic rules\;
    \For{each episode}{
        \For{$t \in T_e$}{
            $\textbf{s}_g \leftarrow$ \text{Graph representation of state} $s$\;
            $\textbf{a}\leftarrow$ \text{Categorical over $P_a$ and action sampling}\;
            \If{$t \mod \tau = 0$}{
                \tcp{Periodically reward shaping}
                $\textbf{$r^*$} \leftarrow$ 
                \FReasoner{$s_g$, $r$, \text{explainer\_obj}}\;
            }
        }
        \tcp{Generate Trajectory \(\mathcal{T} = \{s_t, a_t, r_{t+1}\}_{t=1}^T\) and store log probability of $a$}

         \tcp{Update Policy}
    \tcp{Compute based on $\mathcal{T}$: discounted return $G_t$, mean return $\bar{G}$, std deviation $\sigma_G$, advantage $\delta_t$, and entropy bonus $H_t$}
    \tcp{Update $\theta$ to minimize total loss $\mathcal{L}$}
    $\boldsymbol{\mathcal{L} \leftarrow -\nabla_\theta \log \pi_\theta(s_g, a) \frac{\delta}{\sigma_G} - \lambda_H H}$
    }
   
\end{algorithm}
The update rule based on the policy gradient theorem~\cite{refpolicytheorem} is, 
\begin{equation}
    \theta_{\text{new}} = \theta_{\text{old}} + \alpha \sum_{t=0}^{\infty} \nabla_\theta \log \pi_\theta(a_t|s_t) \cdot G_t,
\end{equation}
where $\alpha$ is the learning rate. A prevalent challenge associated with the REINFORCE algorithm lies in its \emph{high variance}, attributable to its reliance on Monte Carlo estimation techniques. To ameliorate this issue, the refined version of the proposed REINFORCE algorithm incorporates a baseline. This baseline is subtracted from the returns, i.e., advantage acting as a counterbalance to reduce the variance of the updates,
\begin{equation}
    \theta_{\text{new}} = \theta_{\text{old}} + \alpha \sum_{t=0}^{\infty} \nabla_\theta \log \pi_\theta(a_t|s_t) \cdot (G_t - b_{s_t}), 
\end{equation}
where $b_{s_t}$, represents the baseline. Notably, implementing this baseline diverges from the conventional state-dependent value function that is commonly employed. We normalized the advantage via the discounted return, subtracting the mean and dividing by the standard deviation to stabilize the learning. Furthermore, we leverage an \emph{entropy bonus} approach $H = -\sum_{a} p(a) \log p(a)$ into the loss function to promote exploration by the agent, especially during the initial phases of training. It ensures a broad search of the action space and prevents the policy from prematurely settling into a fixed and deterministic behavior. The loss is given by,
\begin{equation}
    \mathcal{L} \leftarrow -\nabla_\theta \log \pi_\theta(s_g, a) \frac{\delta}{\sigma_G} - \lambda_H H,
\end{equation}
where $\lambda_H$ is entropy weight and $H$ denotes entropy bonus. This method encourages actions to optimize future rewards using policy loss while maintaining randomness via the entropy bonus for a balanced exploration-exploitation trade-off.

\subsection{GNN-REINFORCE}
\label{Sec:GNN-REINFORCEtech}
As depicted in Figure~\ref{fig:TANGO_workflow}, we propose an approach to \emph{transform} the state space of conventional network environments into a graph format. By parsing the features of a state into nodes and mapping their interrelations to edges, our approach provides a more prosperous and intuitive representation. This is especially advantageous in scenarios where relational dynamics are critical to the network environment. Furthermore, the graph-oriented structure of states enables the algorithm to scale more effectively to expansive state spaces compared to conventional approaches. This could represent a significant advancement in making RL techniques more practical for tackling real-world, large-scale challenges with intricate state spaces in 6G networks. The structure of the graph used in the model is outlined in Table~\ref{tab:node_features_edges}. Each node represents a combination of different features from the state space of the GRL agent and shows how the nodes are interconnected.

\begin{table}[t!]
\centering
\caption{Node Features and Edge Relationships Based on the State Space (See Section~\ref{sec:Optimization})}
\begin{tabular}{|c|c|c|c|}
\hline
\textbf{Node Index} & \textbf{Feature 1} & \textbf{Feature 2} & \textbf{Connected To} \\
\hline
0 & \( s_1^t \) (SNR) & \( s_2^t \) & 1, 2, 3 \\
\hline
1 & \( s_3^t \) (Residual Capacity) & \( s_4^t \) & 0, 2, 3 \\
\hline
2 & \( s_1^t \) (SNR) & \( s_4^t \) & 0, 1, 3 \\
\hline
3 & \( s_2^t \) (Traffic Volume) & \( s_3^t \) & 0, 1, 2 \\
\hline
\end{tabular}
\label{tab:node_features_edges}
\end{table} The architecture of the \emph{TANGO} model can be summarized by the
following operations:

\textbf{GCN Layer}: GCNs are a subset of GNNs that draw inspiration from convolutional neural networks (CNNs). They are designed for graph data, updating node features by aggregating neighbor features. Unlike GNNs, GCNs have a simpler structure and include a normalization factor based on the node's degree. GCNs propagate data across a graph and train based on signals or features of a graph $\Tilde{G} = (V, E)$, where  $V$ represents the nodes and $E$  signifies the edges. The method used to spread this information is governed by a propagation rule, 

\begin{equation}
H^{(l+1)} = \tilde{\sigma}\left(\tilde{D}^{-\frac{1}{2}}\tilde{A}^\prime\tilde{D}^{-\frac{1}{2}}\tilde{H}^{(l)}W^{(l)}\right),
\end{equation}
where at layer $l$, $\tilde{H}^{(l)}$ represents the node state matrix and each row pertains to a node. The adjusted adjacency matrix $\tilde{A}^\prime$ is obtained by adding the diagonal matrix $\tilde{I}$ to $\tilde{A}$. The degree matrix for $\tilde{A}^\prime$, denoted as $\tilde{D}$.  A learnable matrix of weights is represented by $W^{(l)}$, while  $\tilde{\sigma}$ denotes an activation function (e.g., rectified linear unit (ReLU)).

\textbf{Layer Normalization}: In a GCN layer, node features evolve by aggregating data from neighboring nodes. To ensure stability during training, these features must not exhibit drastic variations. Layer normalization stabilizes training by efficiently scaling and centering these features, potentially accelerating the learning process~\cite{REINFORCECai}.


\textbf{Pooling}: GNNs often necessitate the transformation of graphs into fixed-size representations, which might vary in node count. The \emph{global add pool (GAP)} technique emerges as an effective solution, summarizing node features to generate a uniform graph-level feature vector~\cite{gnnpoolingref}. Associated with graph $\tilde{G}$ is a node feature matrix $X$ with dimensions $|V| \times F$ where $|V|$ represents the total number of nodes in $\tilde{G}$ and $F$ indicates the feature dimensions associated with each node. The GAP operation can be mathematically expressed as,

\begin{equation}
    x_{\tilde{g}} = \sum_{j=1}^{|V|} x_j,
\end{equation}
where, $x_{\tilde{g}}$ is the graph-level feature vector with a consistent length of $F$  and $x_j$ signifies the feature vector of the $j$-th node. GAP provides a comprehensive, fixed-dimensional graph representation by aggregating individual nodes' features. 

\textbf{Linear Layers and Dropout}: The given architecture details two consecutive dense layers within a neural model. Dense layers, or fully connected layers, ensure each neuron from the input communicates with every output neuron. A weight matrix, alongside a bias vector, determines their connection. A scaled exponential linear unit (SELU) function is applied post the first layer,
\begin{equation}
\text{SELU}(x) = \tilde{\lambda} \left( x \cdot \mathbbm{1}_{(x > 0)} + \tilde{\alpha} (e^x - 1) \cdot \mathbbm{1}_{(x \leq 0)} \right),
\end{equation}
where the parameters $\tilde{\lambda}$ and $\tilde{\alpha}$ are predefined  scaling coefficients. Experimentally, we have observed that the SELU and Alpha Dropout~\cite{REINFORCEGuo} combination exhibits remarkable performance. Furthermore, acknowledging the dynamic nature of the 6G network environment, we integrate a \emph{learning rate scheduler}~\cite{REINFORCEXiong} into the framework. It allows for the adaptive adjustment of the learning rate throughout the training process. This method contributes to enhancing training stability and promotes more efficient convergence~\cite{reflearningrateXu}.


\textbf{Readout}: In GNNs, the readout phase involves creating a unified representation of the entire graph by applying a function to the individual representations of all nodes. This function, which can be trained to improve performance, can include methods such as summing node representations, averaging them, or using advanced techniques such as attention mechanisms to determine each node's importance. The resulting comprehensive graph representation contains the complete graph information. It can be used as input for subsequent models, such as a multilayer perceptron (MLP), to perform tasks like classification or regression for final predictions. This phase is essential for generating a vector that captures the graph's key features and structure, enabling effective decision-making based on the entire graph rather than individual nodes.

\subsection{Symbolic Subsystem}
\subsubsection{TANGO Explainer}
\label{Sec:TANGO_Explainer}
One of the primary challenges in modern statistics is the approximation of complex probability densities. Variational inference (VI)~\cite{baysinfJordan}, also called~\cite{baysinfKonishi, baysinfBlei, baysinfZhang} variational Bayes (VB), or variational approximation (VA), is an optimization-based technique for approximate Bayesian inference. This method provides a computationally efficient alternative to sampling methods~\cite{baysinfTran} and it can be used to estimate the conditional density of latent variables using observed variables~\cite{baysinfKucukelbirb}. In this context, the proposed Bayesian GNN explainer approach leverages a dual advantage. First, it allows introspection into the GRL agent's decision-making process, highlighting the critical components of the input graph (state) that significantly influence decisions. Moreover, it can quantify the uncertainty of these explanations, giving more profound insights into their reliability.

More specifically, the VI method used in \emph{Explainer} incorporates latent variables ($\mathbf{z}$), such as node $\mathcal{I}_{x}$ and edge $\mathcal{I}_{i}$ masks, which we aim to infer, along with observed data ($\mathbf{x}$), which includes the topological structure of the graph and its concomitant features. This method involves the development of a probabilistic model for both the latent and observed variables, represented as $p(z, x)$ and referred to as the posterior. By applying Bayes' theorem, we can calculate the posterior probability density as,
\begin{equation}
\underbrace{p(\mathbf{z} \mid \mathbf{x})}_{\text{Posterior}} = \underbrace{p(\mathbf{z})}_{\text{Prior}} \times \frac{\overbrace{p(\mathbf{x} \mid \mathbf{z})}^{\text{Likelihood}}}{\underbrace{\int p(\mathbf{x} \mid \mathbf{z}) \, p(\mathbf{z}) \, \mathrm{d}\mathbf{z}}_{\text{Marginal Likelihood (Evidence)}}} \enspace ,
\end{equation}
where the prior probability, denoted as p($\mathbf{z}$), represents the initial beliefs regarding the distribution of $\mathbf{z}$ before incorporating any new evidence or data. The probability $p(\mathbf{x} \mid \mathbf{z})$ represents how likely the observed data is given a specific configuration of the hidden variables. The marginal likelihood p($\mathbf{x}$) can be regarded as the integral of the likelihood over the hidden variables, which is expressed as $p(\mathbf{x}) =  {\int p(\mathbf{x}, \mathbf{z})\mathrm{d}\mathbf{z}}= {\int p(\mathbf{x} \mid \mathbf{z}) \, p(\mathbf{z}) \, \mathrm{d}\mathbf{z}}$.
VB aims to find the best approximation of the posterior distribution by a probability distribution that belongs to a family of tractable distributions, denoted as $q^\star(\mathbf{z})$,
\begin{equation}
q^\star(\mathbf{z}) = \underbrace{\text{argmin}}_{q(\mathbf{z}) \in \mathrm{Q}} \text{ KL}\left(q(\mathbf{z}) \, \lvert\lvert \, p(\mathbf{z} \mid \mathbf{x}) \right) \enspace ,
\end{equation}
where the Kullback-Leibler divergence is denoted by $\text{KL}(. \lvert \lvert.)$,
\begin{equation}
\text{KL}\left(q(\mathbf{z}) \, \lvert\lvert \, p(\mathbf{z} \mid \mathbf{x}) \right) = \int q(\mathbf{z}) \, \text{log } \frac{q(\mathbf{z})}{p(\mathbf{z} \mid \mathbf{x})} \mathrm{d}\mathbf{z} \enspace.
\end{equation}

The KL divergence measures the difference between two probability distributions, specifically q($\mathbf{z}$) and $p(\mathbf{z} \mid \mathbf{x})$. It provides a way to assess the closeness or divergence of these two distributions in VI method,
\begin{subequations}
\begin{align}
KL(q(z) \| p(z|x)) &= \int q(z) \log \frac{q(z)}{p(z|x)} \, dz \\
&= - \int q(z) \log \frac{p(z|x)}{q(z)} \, dz \\
&= - \left( \int q(z) \log \frac{p(x, z)}{q(z)} \, dz \right. \nonumber \\
&\qquad \left. - \int q(z) \log p(x) \, dz \right) \\
&= - \int q(z) \log \frac{p(x, z)}{q(z)} \, dz \nonumber \\
&\qquad + \log p(x) \int q(z) \, dz \\
&= - \mathcal{L}_{ELBO}(q) + \log p(x) \label{eqn:elbo}
\end{align}
\end{subequations}
where \(\mathcal{L}_{ELBO}(q)\) is referred to as the variational lower bound or evidence lower bound (ELBO).  Equation~\ref{eqn:elbo} arises from the fact that \(\int q(z) dz = 1\). Upon further dissection of ELBO,
\vspace{-.3cm}
\begin{subequations}
\begin{align}
\mathcal{L}_{ELBO}(q) &= \int q(z) \log \frac{p(x, z)}{q(z)} \, dz \\
&= \int q(z) \left( \log p(x, z) - \log q(z) \right) \, dz \\
&= \int q(z) \log p(x, z) \, dz - \int q(z) \log q(z) \, dz \\
&= \mathbb{E}_q [\log p(x, z)] + \mathcal{H}(q) \label{eqn:elbo_final}\\
&= \mathbb{E}_q [\log p(x, z)] - \mathbb{E}_q [\log q(z)]
\end{align}
\end{subequations}
where the entropy of \(q\) is denoted as \(\mathcal{H}(q) = -\int q(z) \log q(z) \, dz\). The initial component in Equation~\ref{eqn:elbo_final} is a form of energy that prompts \(q(z)\) to concentrate the probability distribution where the model indicates high likelihood \(p(x, z)\). The second component, entropy, promotes \(q(z)\) to disperse throughout the space and prevent concentration in a single location. 
The marginal likelihood is also referred to as evidence, and it is evident that the ELBO indeed represents a lower bound for the evidence, while KL divergence is non-negative,
\vspace{-.2cm}
\begin{subequations}
\begin{align}
\mathcal{L}_{ELBO}(q) &= -\left(\text{KL}\left(q(\mathbf{z}) \,\middle\|\, p(\mathbf{z} \mid \mathbf{x}) \right) - \text{log } p(\mathbf{x})\right) \\
\text{log } p(\mathbf{x}) &= \mathcal{L}_{ELBO}(q) + \text{KL}\left(q(\mathbf{z}) \,\middle\|\, p(\mathbf{z} \mid \mathbf{x}) \right) \\
\text{log } p(\mathbf{x}) &\geq \mathcal{L}_{ELBO}(q).
\end{align}
\end{subequations}
Minimizing $\text{KL}\left(q(\mathbf{z}) \, \lvert\lvert \, p(\mathbf{z} \mid \mathbf{x}) \right)$ results in the same outcome as maximizing $\mathcal{L}(q)$,

\begin{subequations}
\begin{align}
q^\star(\mathbf{z}) &= \underbrace{\text{argmin}}_{q(\mathbf{z}) \in \mathrm{Q}} \text{ KL}\left(q(\mathbf{z}) \,\middle\|\, p(\mathbf{z} \mid \mathbf{x}) \right), \\[0.5em]
q^\star(\mathbf{z}) &= \underbrace{\text{argmax}}_{q(\mathbf{z}) \in \mathrm{Q}} \mathcal{L}(q), \\[0.5em]
q^\star(\mathbf{z}) &= \underbrace{\text{argmax}}_{q(\mathbf{z}) \in \mathrm{Q}} \left\{\mathbb{E}_q [\log p(x, z)] -\mathbb{E}_q[\log q(z)] \right\}.
\end{align}
\end{subequations}

One widely recognized variational method is the mean-field approximation framework, representing a factorized form of VI~\cite{bookmeanBishop}. In mean field variational Bayes (MFVB), we posit that the variational distribution can be deconstructed into independent distributions over separate groups ($m$ is the total number of such groups) of latent variables \(z_j\) for \(j = \{1, \ldots, m\}\), which are mutually independent and each influenced by its variational density,
\[
q(\mathbf{z}) = \prod_{j=1}^m q_j(z_j).
\]
This factorization simplifies the calculation of the ELBO, as it reduces the complexity of the required integration. The ELBO for MFVB is then given by,
\begin{subequations}
\begin{align}
\mathcal{L}_{ELBO}(q) &= \mathbb{E}_{q(\mathbf{z})}\left[\text{log } p(\mathbf{z}, \mathbf{x})\right] -  \mathbb{E}_{q(\mathbf{z})}\left[\text{log } q(\mathbf{z}) \right] \\[.5em]
&= \int \prod_{i=1}^m q_i(z_i) \, \text{log } p(\mathbf{z}, \mathbf{x}) \, \mathrm{d}\mathbf{z} \nonumber \\
&\quad - \int \prod_{i=1}^m q_i(z_i) \, \text{log}\prod_{i=1}^m q_i(z_i) \, \mathrm{d}\mathbf{z}.
\end{align}
\end{subequations}

In the subsequent steps, we optimize ELBO by concentrating on a singular variational density \(q_j(z_j)\) while maintaining the constancy of all others,

\begin{subequations}
\begin{align}
\mathcal{L}_{ELBO}(q_j) &= \int \prod_{i=1}^m q_i(z_i) \, \text{log } p(\mathbf{z}, \mathbf{x}) \, \mathrm{d}\mathbf{z} \nonumber \\
&\quad - \int \prod_{i=1}^m q_i(z_i) \, \text{log}\prod_{i=1}^m q_i(z_i) \, \mathrm{d}\mathbf{z} \\[.5em]
&= \int q_j(z_j) \, \mathbb{E}_{q(\mathbf{z}_{-j})}\left[\text{log } p(\mathbf{z}, \mathbf{x})\right]\mathrm{d}z_j \nonumber \\
&\quad - \int q_j(z_j) \, \text{log } q_j(z_j) \, \mathrm{d}z_j.
\end{align}
\end{subequations}

By introducing $\tilde{p}{(\mathbf{x}, z_j)}$, we facilitate the derivation of the optimal form for each factor $q_j^\star(z_j)$~\cite{bookmeanBishop} within the mean-field approximation,
\[
\text{log } \tilde{p}{(\mathbf{x}, z_j)} = \mathbb{E}_{q(\mathbf{z}_{-j})}\left[\text{log } p(\mathbf{z}, \mathbf{x})\right] - \phi,
\]
where $\phi$ is the log normalizing constant ensuring that $\tilde{p}{(\mathbf{x}, z_j)}$ integrates to one. The ELBO can be expressed in a way that involves $\tilde{p}{(\mathbf{x}, z_j)}$,
\begin{align}
\mathcal{L}_{ELBO}(q_j) &\propto \int q_j(z_j) \, \text{log } \tilde{p}(\mathbf{x}, z_j) \, \mathrm{d}z_j \nonumber \\
&\quad - \int q_j(z_j) \, \text{log } q_j(z_j) \, \mathrm{d}z_j \\[.5em]
&= \int q_j(z_j) \, \text{log } \frac{\tilde{p}(\mathbf{x}, z_j)}{q_j(z_j)} \, \mathrm{d}z_j \\[.5em]
&= -\int q_j(z_j) \, \text{log } \frac{q_j(z_j)}{\tilde{p}(\mathbf{x}, z_j)} \, \mathrm{d}z_j \\[.5em]
&= -\text{KL}\left(q_j(z_j) \, \lvert\lvert \, \tilde{p}(\mathbf{x}, z_j)\right).
\end{align}
The ELBO is maximized when the KL divergence between the variational distribution \(q_j(z_j)\) and \(\tilde{p}(\mathbf{x}, z_j)\) is minimized. The KL divergence is zero when the variational distribution \(q_j(z_j)\) equals \(\tilde{p}(\mathbf{x}, z_j)\). The optimal variational density \(q_j^\star(z_j)\) can be derived as,

\begin{equation}
\begin{aligned}
q_j^\star(z_j) &= \text{exp}\left(\mathbb{E}_{q_{-j}(\mathbf{z}_{-j})}\left[\text{log } p(\mathbf{x}, \mathbf{z}) \right] - \phi\right) \\[.5em]
&= \frac{\text{exp}\left(\mathbb{E}_{q_{-j}(\mathbf{z}_{-j})}\left[\text{log } p(\mathbf{x}, \mathbf{z}) \right]\right)}{\int \text{exp}\left(\mathbb{E}_{q_{-j}(\mathbf{z}_{-j})}\left[\text{log } p(\mathbf{x}, \mathbf{z}) \right]\right) \mathrm{d}z_j}.
\end{aligned}
\end{equation}
This approach does not offer a straightforward solution as each optimal variational density depends on the others. Therefore, an iterative method is necessary. The process begins by initializing all factors \(q_j(z_i)\) and then iteratively updating each one, utilizing the latest updates of the others. This technique is known as coordinate ascent variational inference (CAVI)~\cite{bookmeanBishop, meanNguyen, meanBhattacharya}. The CAVI algorithm is based on the previously mentioned equations, which are outlined in Algorithm~\ref{algo:TANGO_GRL_CAVI}~\cite{baysinfBlei}. A Python code example for performing coordinate ascent to optimize mask parameters in TANGO is available online\footnote{\url{https://github.com/frezazadeh/papers/blob/main/TANGO/Coordinate_Ascent_for_Mask_Parameter.py}}.

\begin{algorithm}[ht]
    \caption{Coordinate ascent variational inference}
    \label{algo:TANGO_GRL_CAVI}
    \small
    \SetKwFunction{FMain}{REINFORCE}
    \SetKwFunction{FExplainer}{Explainer}
    \SetKwFunction{FReasoner}{Reasoner}
    \SetKwInput{KwRequire}{Require}
    \SetKwInput{KwInput}{Input}
    \SetKwInput{KwOutput}{Output}
    \SetKwInput{KwInitialize}{Initialize}
    \KwInput{A model $p(z, x)$, a data set $x$}
    \KwOutput{A variational density $q(z) = \prod_{j=1}^m q_j(z_j)$}
    \KwInitialize{Variational factors $q_j(z_j)$}
    \While{the ELBO has not converged}{
        \For{$j \in \{1, \ldots, m\}$}{
            Set $q_j(z_j) \propto \exp\{\mathbb{E}_{-j}[ \log p(z_j | z_{-j}, x)]\}$\;
        }
        Compute ELBO($q$) = $\mathbb{E}[\log p(z, x)] + \mathbb{E}[\log q(z)]$\;
    }
    \Return{$q(z)$}
\end{algorithm}

The Bayesian GNN explainer employs VI to approximate the posterior distribution of the edges and nodes masks, which indicates the importance of each edge and node feature in the graph using observed data. The explainer outputs the node and edge masks, $\mathcal{I}_{x}$ and $\mathcal{I}_{i}$, respectively, along with their associated uncertainties. Notice that in mathematical and conceptual terms, the standard deviation represents the idea of uncertainty in the Gaussian distribution~\cite{Xiaotong_uncer_ref}.

\vspace{-.5cm}
\subsubsection{TANGO Reasoner}\label{Sec:TANGORewardShaper}
The proposed methodology introduces a novel approach to symbolic reasoning by leveraging a custom-developed symbolic reasoner. This reasoner integrates explainer outputs with graph features to derive actionable insights. Functioning on a rule-based system, the reasoner allows for dynamic evaluation and adjustment of nodes within a network. The symbolic reasoning module is initialized with the flexibility to manage and execute predefined rules and dynamically incorporate new rules. Rules serve as the foundation of the reasoning process, each comprising a condition and an action.

The condition is a logical~\cite{Seo_rule_ref} expression that evaluates the state of a node based on its attributes, such as importance, uncertainty, and similarity. These rules collectively form a comprehensive rule base. For each node $v$ in the set of nodes $V$, it operates on the node features $s_x$ and edge index $s_i$. Node is considered highly important if its importance (i.e., feature importance $\mathcal{I}_{x}^v$ and importance of the connecting edges $\mathcal{I}_{i}^v$) surpasses predefined thresholds ($\mathcal{I}^v_{\varrho}$, $\mathcal{I}^i_{\varrho}$) and its uncertainties fall below \(\mathcal{U}_{\varrho}\) threshold. This categorization assists in identifying nodes that are both important and reliable. The cosine similarity $\mathcal{Z}$ is calculated between the features of linked nodes to assess the strength of their connections. Nodes with strong connections (similarity exceeding a specific threshold, i.e.,  $\text{cosine similarity} > \mathcal{Z}_{\varrho}$) are specified, as these robust connections can impact the reward shaping. We consider two particular rules: firstly, nodes showing high importance, low uncertainty, and strong relationship similarity are positively adjusted to promote significant and reliable nodes. Secondly, nodes with high importance but low similarity receive a penalty, aiding in identifying and mitigating isolated yet important nodes.

The module systematically applies each rule to the nodes, checking the conditions against the current state of each node. If a condition is satisfied, the associated action is performed, and the result is recorded. The adjustments for each node are determined by aggregating the results of all applicable rules. The framework then sums the outcomes of the rule applications to compute a final adjustment value for reward shaping in GRL. The reward shaping process utilizes the symbolic reasoner to adjust rewards based on the evaluated metrics between nodes dynamically. The culminating reward is the  summation of the base reward $\mathcal{R}$, and XAI reward $r^*$,

\begin{equation}
    \mathcal{R}_{Final} = \mathcal{R} + r^* .
\end{equation}

This approach paves the way for extracting symbolic knowledge. The method bridges the gap between abstract neural network patterns and symbolic knowledge by translating the variational posterior distributions of the edge and node feature masks into interpretable insights. Indeed, the edge and node feature masks act as a bridge, translating the abstract, high-dimensional patterns learned by the neural network into a form that is analogous to symbolic knowledge. This translation and symbolic representation elucidate the specific graph elements--edges and nodes--most influential in the network's predictions, akin to identifying key variables and rules in a symbolic system. The symbolic representation of the neural network's decision-making enhances interpretability. It relates interpretations to specific components of the input graph, offering a robust tool for explicating the complex mechanisms of neural network-based decision-making. 

\section{Performance Evaluation}
\label{sec:perf_eval}
In this section, we evaluate our proposed framework in the decision engine of the testbed on a dedicated server equipped with two Intel(R) Xeon(R) Gold 5218 CPUs @ 2.30GHz and two NVIDIA GeForce RTX 2080 Ti GPUs. The model is implemented in PyTorch and PyTorch Geometric (PyG)\footnote{\url{https://pytorch-geometric.readthedocs.io/en/latest/}.}, widely used deep learning frameworks. Our TANGO model, integrates GCN with REINFORCE algorithm principles. The detailed architecture of the neural network is comprehensively discussed in Section~\ref{Sec:GNN-REINFORCEtech}. We comprehensively examine AI, complexity, energy consumption, robustness, network performance, scalability, and explainability metrics to ensure a thorough analysis of the proposed algorithm's performance.

\textbf{Hyperparameter}: Delving into reasoner parameters, these are calculated based on the output of the explainer to modify the reward function to enhance learning efficiency. The \emph{Importance Thresholds}~\(\mathcal{I}^v_{\varrho}\) and \(\mathcal{I}^i_{\varrho}\) establish a benchmark for assessing feature and node importance within the graph, which are set to 0.8. The \emph{Penalty} is a negative reward applied under specific conditions to direct the agent's learning process, valued at -1. The \emph{Uncertainty Threshold}~\(\mathcal{U}_{\varrho}\) sets the level of uncertainty below which a node or feature is deemed important, established at 0.2. The \emph{Relationship Strength Threshold}~ \(\mathcal{Z}_{\varrho}\) determines the minimum level of inter-node relationship strength required to be considered significant, set at 0.8. Moreover, the \emph{Alpha Dropout Probability} dictates the dropout rate during training, a method employed to prevent overfitting, with a value of 0.1. The \emph{Entropy Weight} is utilized in the REINFORCE algorithm to boost exploration, set at 0.01. The \emph{Step Size} and the \emph{Gamma} are parameters of the learning rate scheduler, which optimally adjusts the learning rate during training to enhance convergence, with values of 100 and 0.1, respectively.

\textbf{Training phase vs. Inference phase}: In DRL, the training phase and inference phase serve different purposes. During training, the algorithm interacts with the environment, learns from these interactions, and adjusts its policy or value functions based on the rewards it receives. This involves continuous improvement through trial and error, often using techniques such as Q-learning, policy gradients, or actor-critic methods to maximize total rewards. In contrast, the inference phase uses the trained model to make decisions in a real-world or simulated environment without further learning. The model applies the policy or value functions learned during training to select actions and aims to achieve optimal performance based on prior learning. The key distinction is that training focuses on learning and improving the model, while inference focuses on applying the learned model to make effective decisions.

\textbf{Traffic generating}: As shown in Figure~\ref{fig:testbed_architecture}, MGEN (Multi-Generator) is a versatile software tool developed by the network research group at the Naval Research Laboratory~\footnote{\url{https://github.com/USNavalResearchLaboratory/mgen}}. Its purpose is to generate real-time traffic patterns for assessing network performance. MGEN can handle both UDP and TCP traffic, accommodating a wide range of testing scenarios. The tool utilizes script-based configurations~\cite{MGENZanzi} to define traffic flows, enabling the specification of parameters such as packet size, inter-packet intervals, and traffic duration. By executing these scripts, MGEN generates traffic that can be observed and analyzed, offering valuable insights into network performance, capacity, and protocol efficiency under diverse conditions. 



\textbf{Cell configuration}: Table~\ref{table:cells} provides details of the multi-cell configuration for the Callbox Ultimate and Callbox Mini setups using Amarisoft, specifically focusing on the 5G New Radio (NR) technology within band n7. It includes four cells, with the first three cells designated as part of the Ultimate Callbox (Ultim.) and the fourth cell as part of the Mini Box (Mini). All cells operate on band n7 (i.e., the downlink frequency range of 2620 to 2690 MHz), with specific downlink absolute radio frequency channel numbers (ARFCN) and synchronization signal block (SSB) ARFCN provided for precise frequency tuning. The setup maintains a consistent bandwidth of 10 MHz and a subcarrier spacing of 15 kHz~\cite{Abdis_ref} across all cells, ensuring uniformity in channel spacing and frequency allocation. This analysis concentrates only on the first cell, which handles all downlink traffic. According to the configurations, it is important to note that each cell supports 52 PRBs, which is directly correlated with action space $\mathcal{A}_{\text{max}}$. We set \(\Delta_{\text{over}}\) to -2 and \(\Delta_{\text{under}}\) to 2, which establish the bounds for over-provisioning and under-provisioning, respectively.
\begin{table}[ht]
\vspace{-.3cm}
\caption{Multi-cell Configuration}
\label{table:cells}
\resizebox{\columnwidth}{!}{%
\begin{tabular}{|c|c|c|c|c|c|c|c|}
\hline
cell & gNB  & band & type & dl arfcn & ssb arfcn & bandwidth & sub. spacing \\ \hline
1    & Ultim. & n7   & NR   & 525000        & 524910         & 10 MHz    & 15 kHz              \\ \hline
2    & Ultim. & n7   & NR   & 528000        & 528030         & 10 MHz    & 15 kHz              \\ \hline
3    & Ultim. & n7   & NR   & 531000        & 530910         & 10 MHz    & 15 kHz              \\ \hline
4    & Mini  & n7   & NR   & 525000        & 524910         & 10 MHz    & 15 kHz              \\ \hline
\end{tabular}%
}
\end{table}

\textbf{Baseline and Benchmarks}: In this study, the baseline used is the GNN-REINFORCE algorithm, also referred to as TANGO, without the inclusion of the symbolic subsystem. This work employs comparison benchmarks, including Dueling DQN~\cite{UDQNWang}, Double DQN~\cite{DDQNHasselt, Specialization_TVT}, DQN, and REINFORCE~\cite{REINFORCEZhang}. Dueling DQN improves upon the standard DQN by separately estimating the state value and advantage functions, resulting in enhanced learning efficiency and stability. Double DQN addresses the overestimation bias of DQN by separating action selection from value evaluation in Q-learning updates. The traditional DQN employs a neural network to estimate Q-values for action-state pairs and learns optimal policies through experience replay and a target network for training stabilization. On the other hand, REINFORCE is a policy gradient method that directly optimizes the policy by updating it to increase expected rewards using sampled trajectories.

\subsection{AI Metric}
\label{sec:ai_met}
\begin{figure}[t!]
\centering
\includegraphics[width=.9\columnwidth]{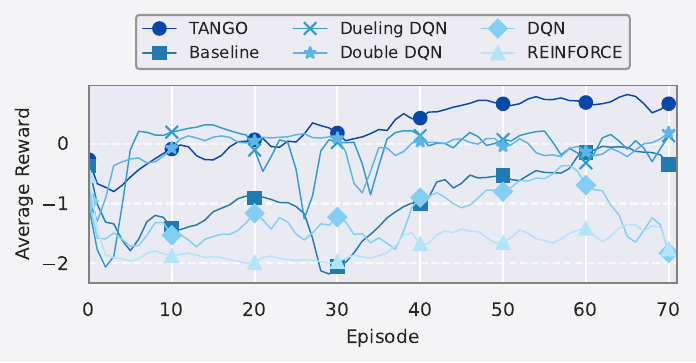}
\caption{The graph compares TANGO's average reward trajectories per episode to Baseline (without symbolic subsystem) and other DRL benchmarks during 70 episodes. The curves are smoothed (Gaussian smoothing\cite{Gaussianfilterref}) for visual clarity.}
\label{Sec:reward-training}
\vspace{-.6cm}
\end{figure}

Figure~\ref{Sec:reward-training} shows a comparative analysis of the average reward performance of various algorithms across a sequence of episodes. Upon examination, it is observed that TANGO consistently demonstrates high average rewards, outperforming the Baseline algorithm throughout the observed episodes. There is a significant spike in TANGO's performance around episode 30, following which the algorithm maintains its lead, further suggesting that it has found an efficient way to explore the environment or a superior policy. On the other hand, the Baseline algorithm shows fluctuating performance with a broader range of average rewards, indicating potential instability or inefficiency in learning or exploration. Although the Baseline algorithm occasionally reaches comparable rewards to TANGO, it fails to sustain or improve upon these highs, further highlighting its inconsistency. The benchmark algorithms exhibit varied performances.

Dueling DQN's performance sometimes matched or exceeded that of TANGO, which was particularly noticeable in certain episodes where it spiked. However, these instances were inconsistent, with the average reward often dropping significantly in subsequent episodes. Double DQN displayed a trend of average rewards that were generally below that of TANGO and less volatile than Dueling DQN. Despite some episodes where its performance closely approached TANGO's, it did not sustain these peaks. However, Double DQN did not consistently outperform TANGO in this particular evaluation. DQN's average reward occasionally dipped significantly below that of TANGO, especially notable in the early episodes (before episode 20) and then again around episode 50. These dips indicate that the algorithm was less stable or that it struggled to maintain a policy that consistently maximized rewards. REINFORCE showed an interesting pattern with a sharp fluctuation in average rewards, indicating a high variance in its performance. Like DQN, it also exhibited the potential for high rewards but did not maintain them consistently. However, REINFORCE's rewards were generally lower than TANGO's, particularly in the later episodes, suggesting that it did not learn as effectively as TANGO over time. 

Overall, TANGO is a more effective learner, achieving higher and more stable rewards compared to the Baseline and other benchmark algorithms. This suggests that TANGO has a robust learning approach, which can maintain stable performance even as the learning conditions change throughout the episodes.

\begin{figure}[t!]
\centering
\includegraphics[width=.9\columnwidth]{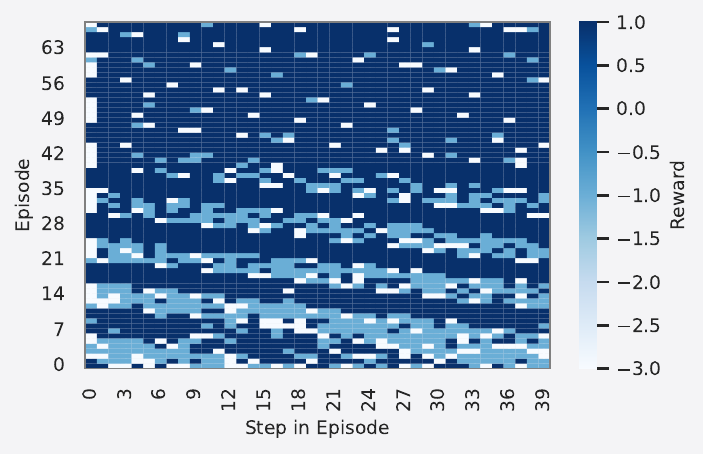}
\caption{The heatmap provides a visual representation of the TANGO reward pattern throughout the episodes in the training phase. Each row is an episode, and each column represents a step within that episode. The color intensity indicates the amount of reward earned at each step.}
\label{Sec:Tango-step-reward-training}

\end{figure}

Figure~\ref{Sec:Tango-step-reward-training} features a heatmap that visually represents the rewards granted to the TANGO algorithm during its training episodes. Each row within the heatmap represents an episode, while each column indicates a step within that episode. The color intensity in the heatmap corresponds to the amount of reward that the algorithm received at each step, with darker blues indicating higher rewards and lighter colors representing lower rewards. The algorithm used a less effective strategy in the early episodes (0 to around 20), resulting in lower rewards. This is evident from the presence of many lighter colors, especially in the beginning steps of each episode. As the episodes progressed into mid-episodes (around 20 to 40), we began to notice more occurrences of darker blues mixed with lighter colors. This suggests that the algorithm is starting to learn strategies that yield more rewards, but its performance still needs to improve. The learning process has begun, and the algorithm is enhancing its policy to maximize the reward. In later episodes (around 40 to 70), dark blue colors become more prevalent, and the lighter colors are less frequent. This indicates that TANGO has learned more effective strategies over time and consistently achieves higher rewards as it navigates the steps within each episode.

\subsection{Complexity Metric}

\subsubsection{Time Complexity} 
\begin{figure}[t!]
\centering
\includegraphics[width=1\columnwidth]{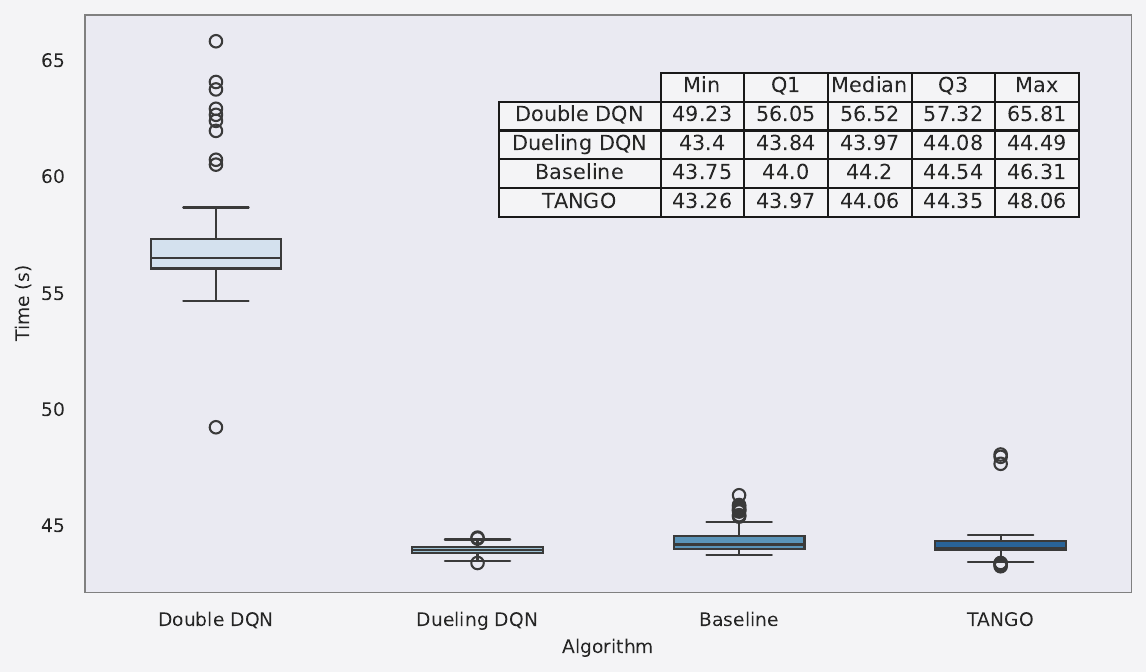}
\caption{Comparative analysis of time complexities for different approaches over a period of 70 episodes during the training phase. The boxplot visually represents the distribution of total computation time per episode for each algorithm. 
}
\label{Sec:Time-training}
\vspace{-.6cm}
\end{figure}

The time it takes to solve a problem based on its size is known as an algorithm's time complexity. Several factors, such as the size of the state and action spaces, the architecture of the neural networks, the number of episodes and steps, and the particular operations of the algorithm, can influence the time complexity of DRL algorithms. As depicted in Figure~\ref{Sec:Time-training}, the present study involves a comparative analysis between the TANGO algorithm and other algorithms, focusing on assessing time complexity over 70 episodes during the training phase. 
From the boxplot, it is clear that TANGO has a tighter interquartile range (IQR) compared to Double DQN and Dueling DQN. The box length in the boxplot depicts the IQR, and TANGO's IQR is from approximately  43.97 s to 44.35 s, which is a narrow range of 0.38 s. This suggests that the TANGO algorithm has less variation in its computation times per episode, indicating a more consistent performance. 

Moreover, TANGO's median time per episode, marked by the line within the box, is slightly less than the Baseline but higher than Dueling DQN. It is significantly lower than Double DQN. This implies that TANGO is generally faster than Double DQN and the Baseline but slightly slower than Dueling DQN. The median time for TANGO is indicated in the table as 44.06 s, close to its first quartile (Q1) of 43.97 s, which shows a skewness towards the lower end of its range. As shown in the table, the maximum value for TANGO is 48.06 s, which is significantly lower compared to Double DQN's 65.81 s, but higher than Dueling DQN's 44.49 s. This implies that while TANGO does not always have the shortest computation time, it avoids the long computation times that can occasionally occur with Double DQN. The presence of outliers is noted for the Double DQN algorithm, as indicated by the points above the upper whisker in the boxplot. TANGO, in contrast, shows no such outliers, which suggests that it maintains a consistent performance without episodes of extreme computation time.

Overall, TANGO offers a favorable balance between efficiency and consistency when compared to Double DQN and Dueling DQN. Despite TANGO being marginally faster on average than the Baseline, as indicated by the slightly lower median time, its superior performance could justify the extra time spent per episode. 

\subsubsection{Computational Complexity }
\begin{figure}[t!]
\centering
\includegraphics[width=1\columnwidth]{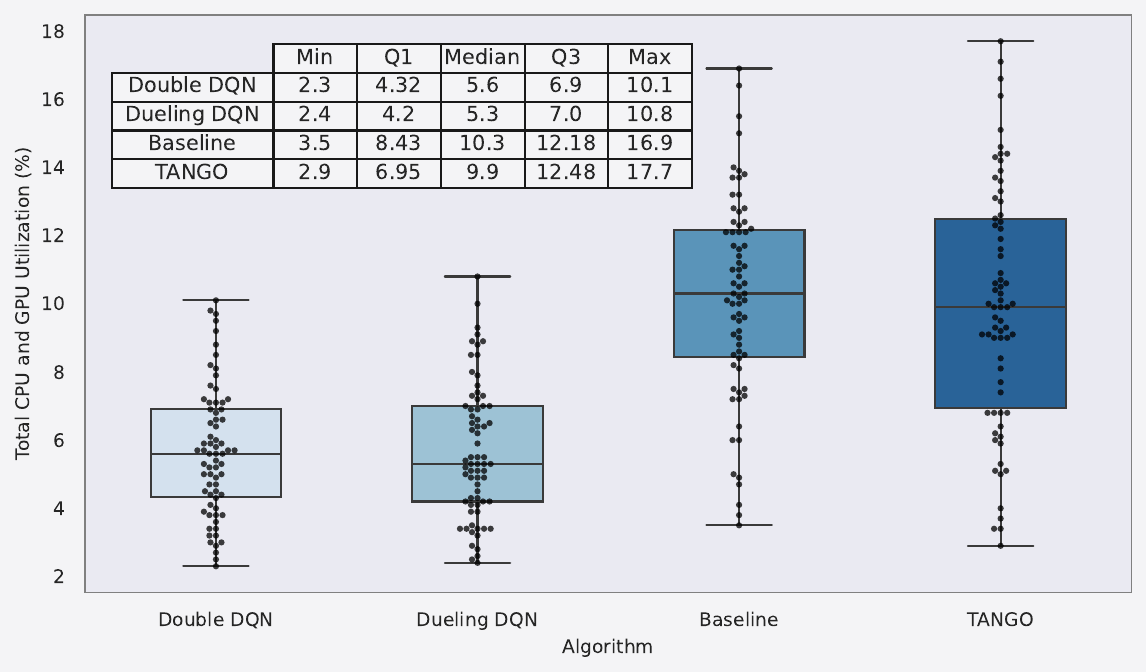}
\caption{Beeswarm boxplot demonstrates the total CPU and GPU utilization during training for each episode. Each point represents the combined CPU and GPU usage for a single episode, while the boxplot overlay summarizes the distribution.
}
\label{Sec:Computtion-training}
\vspace{-.7cm}
\end{figure}

Figure~\ref{Sec:Computtion-training} depicts a beeswarm boxplot that effectively provides a comprehensive understanding of algorithms' resource requirements in terms of the total CPU and GPU utilization during the training phase. TANGO demonstrates a minimum utilization of 2.9\%, which is slightly lower than the Baseline's 3.5\%. However, TANGO's minimum utilization is higher than that of both Double DQN and Dueling DQN, which are at 2.3\% and 2.4\%, respectively. This finding suggests that TANGO's resource usage does not drop as low as the other algorithms, indicating a potentially higher lower bound of resource needs. 

The first quartile for TANGO stands at 6.95\%, which is considerably lower than the Baseline's 8.43\%. However, TANGO's first quartile is higher than both DQN variants. This implies that the lower 25\% of TANGO's resource usage measurements are generally higher than those of the DQN variations but still more efficient than the Baseline. TANGO's median value is 9.9\%, placing it lower than the Baseline at 10.3\%. However, it is significantly higher than both Double DQN and Dueling DQN, which have medians of 5.6\% and 5.3\%, respectively. This means that the middle value of TANGO's resource consumption is less than the Baseline but not as efficient as the DQN variations.
For TANGO, the third quartile is 12.48\%, which is higher than the Baseline's 12.18\% and higher than the DQN variations. This suggests that 75\% of TANGO's resource usage measurements are below 12.48\%, indicating that TANGO generally uses more resources than the Baseline and DQN variations. TANGO's maximum utilization peaks at 17.7\%, which is the highest among all the algorithms compared. This finding suggests that TANGO has episodes where it can consume more resources than any other algorithm, revealing potential spikes in computational demands. The spread of data points (beeswarm aspect of the plot) for TANGO shows a wide distribution of values, with several potential outliers beyond the upper whisker. This shows  variability in its resource utilization, which could imply that TANGO has episodes of training that are significantly more resource-intensive than usual.

\subsection{Energy Metric}
\begin{figure}[t!]
\centering
\includegraphics[width=1\columnwidth]{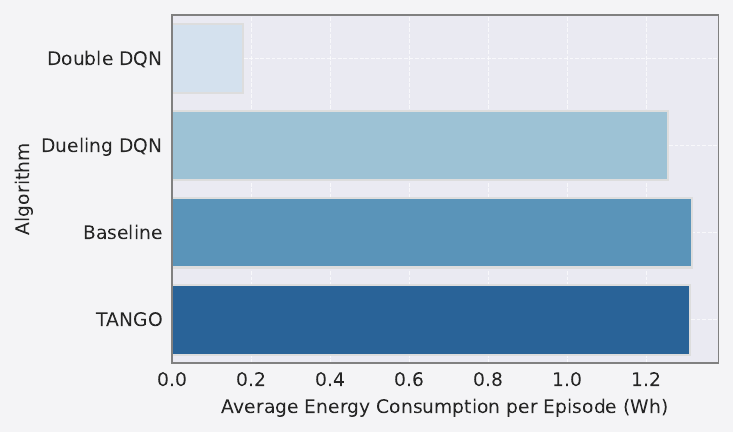}
\caption{The bar chart illustrates the average GPU energy consumption per episode. The results highlight the comparative efficiency of each algorithm in terms of energy usage, providing insights into their performance and sustainability.}
\label{Sec:Energy-training}
\vspace{-.6cm}
\end{figure}

In this experiment, we monitor the GPU's power consumption throughout the algorithms' training process.
According to Figure~\ref{Sec:Energy-training}, TANGO has clear potential optimizations or algorithmic efficiencies (See Section~\ref{Sec:GNN-REINFORCEtech}) that can be used for sustainability, as it consumes slightly less energy than the Baseline algorithm. This improvement in TANGO's energy usage could indicate advancements in algorithm design that contribute to a more sustainable future. While the difference is not substantial, it shows that TANGO is more energy-efficient than the Baseline, albeit by a small margin. TANGO consumes significantly more energy than Dueling DQN. Dueling DQN's specialized architecture that separately computes the value and advantage streams is more energy-efficient while potentially providing an enriched learning mechanism. This comparison suggests that TANGO involves more complex computations, leading to higher energy use. Compared with Double DQN, TANGO's energy consumption is notably higher. TANGO is more energy-efficient than the Baseline, but Dueling DQN and Double DQN outperform it in terms of energy consumption. This implies that TANGO is better suited for scenarios where its potential performance advantages outweigh energy considerations. 

\subsection{Robustness Metric}
\begin{figure}[t!]
\centering
\includegraphics[width=1\columnwidth]{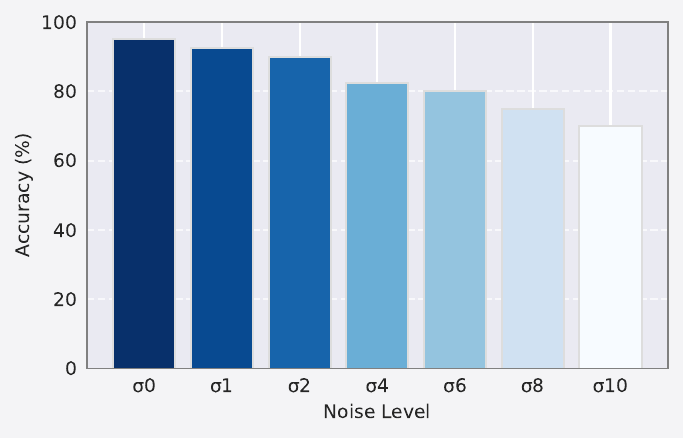}
\caption{Performance analysis of TANGO algorithm under varying noise levels for radio resource allocation, showing robustness across 40 trials in inference phase.}
\label{Sec:ROBUSTVAL1}
\vspace{-.6cm}
\end{figure}

Noise injection~\cite{noiserefFerianc} is a technique used to evaluate the robustness of an AI/ML model--particularly in DRL--by deliberately adding randomness or disturbances to the input data during the inference phase. The main purpose of this approach is to replicate real-world imperfections that the model may not have faced during training. Examples of such imperfections may include signal interference, environmental variations, or other sources of uncertainty that may impact input data quality. By examining the model's reaction to artificially noised inputs, developers can gain a clearer insight into its resilience and ability to sustain performance under less-than-optimal circumstances. Figure~\ref{Sec:ROBUSTVAL1} demonstrates the TANGO algorithm's accuracy in assigning acceptable and optimal PRBs across 40 trials during the inference phase as synthetic noise levels increase. Synthetic noise, modeled as a normally distributed disturbance with a mean of zero and a standard deviation ranging from $\sigma = 0$ (no disturbance) to $\sigma = 10$, is injected into the network states to emulate real-world interference. Remarkably, despite the escalating noise levels, the TANGO algorithm remains resilient, with only a modest decrease in performance. This underscores the algorithm's ability to effectively allocate radio resources even under adverse conditions commonly encountered in communication networks. Here, accuracy refers to the precision with which the GRL agent allocates radio resources in accordance with the specified thresholds for the allocation gap (as explained in the following). Specifically, we set \(\Delta_{\text{over}}\) to -2 and \(\Delta_{\text{under}}\) to~2.

\subsection{Network Metric}
The allocation gap can measure the effectiveness of radio resource allocation algorithms in 5G/6G networks~\cite{allgapLiu, Specialization_TVT}. This refers to the disparity between the intended or preferred distribution of radio frequencies and bandwidth and the actual deployment and utilization of these resources within the network. This gap significantly impacts the overall network performance, particularly in how it handles the amount and type of traffic that the network encounters. The size of the allocation gap is a crucial metric for assessing radio resource allocation algorithms. A smaller gap indicates that the algorithm is adept at forecasting and managing the network's requirements, closely matching the theoretical allocation with the actual usage. Conversely, a larger gap suggests inefficiencies in the network, where the resource allocation is either over-provisioning or under-provisioned. Over-provisioning arises when the network allocates more resources than necessary, which may seem advantageous as it could theoretically handle unexpected spikes in traffic but typically leads to wasteful expenditure of resources, increasing operational costs without a corresponding improvement in user experience. In contrast, under-provisioning occurs when fewer resources than required are allocated, leading to congestion and poor service quality, as the available resources are insufficient to meet user demand.
\begin{figure}[t!]
\centering
\includegraphics[width=1\columnwidth]{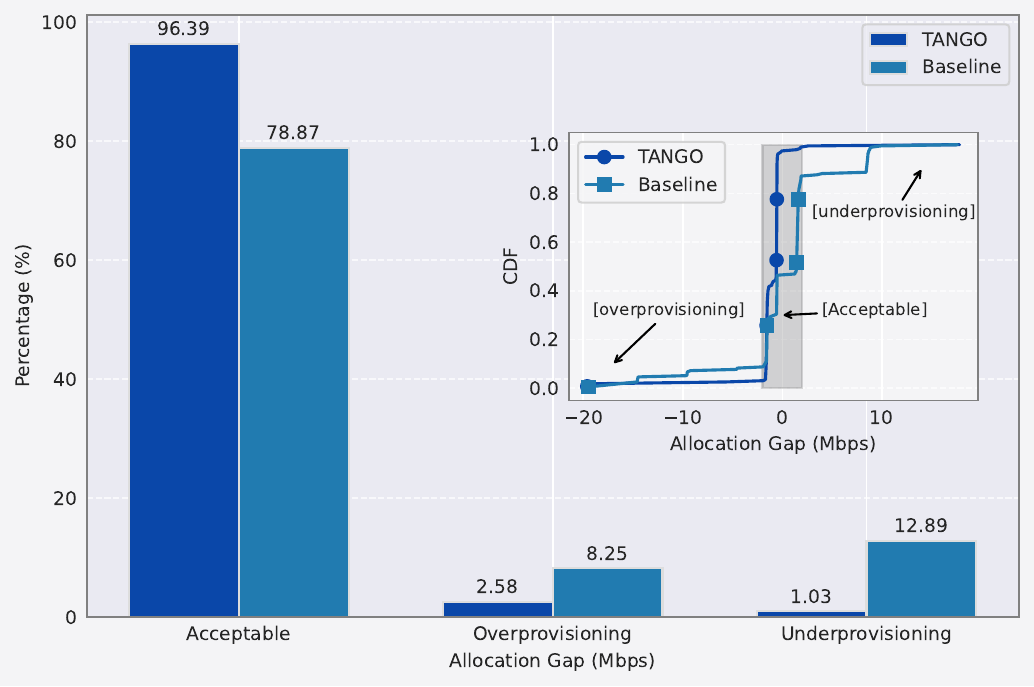}
\caption{The radio resource allocation performance for the TANGO framework and Baseline during the inference phase. The graph provides information about the distribution of allocation gaps, highlighting the zones where acceptable performance is achieved (values within -2 to 2 Mbps). 
}
\label{Sec:Gap_Infernce}
\vspace{-.7cm}
\end{figure}

As shown in Figure~\ref{Sec:Gap_Infernce}, the bar chart and cumulative distribution function (CDF) plots provide quantitative insights into how each method addresses the allocation discrepancies, emphasizing their respective capabilities in managing network resources effectively during the inference phase. It is noteworthy that \emph{the range of -2 to 2 Mbps is defined as a practical tolerance}, considering that traffic demands do not always align perfectly with fixed PRB sizes. This range allows the radio resource allocation algorithm to function within a defined action space, accommodating slight mismatches between PRB-based allocations and the fluctuating traffic demands. The bar chart reveals that the TANGO framework performs significantly better than the Baseline method in terms of achieving optimal allocation, with a high percentage of around 96\%. This suggests that TANGO has a stronger ability to match the actual allocation of radio resources with what is optimally needed. 

Additionally, the comparative sharpness of the CDF plot for TANGO in the acceptable range (-2 to 2 Mbps) implies that TANGO is not only consistent in meeting the targeted allocations but also does so with greater predictive accuracy. This characteristic is essential for dynamic network environments, where traffic patterns can vary significantly, necessitating adaptive and accurate resource allocation strategies. The CDF plot indicates that TANGO manages resource over-provisioning and under-provisioning more effectively than the Baseline method. TANGO's rapid approach to zero in the over-provisioning zone suggests that it rarely allocates more resources than necessary, thereby reducing wastage and potentially lowering operational costs. On the other hand, the Baseline method exhibits a slower decline in this area, indicating a tendency to over-allocate resources, which could lead to inefficiencies and increased costs without proportional benefits. TANGO again shows superior performance in terms of under-provisioning by quickly descending to a lower frequency of under-provisioning instances beyond the acceptable range. This indicates that TANGO is less likely to provide insufficient resources, thereby maintaining service quality and user experience. In contrast, Baseline extends further into higher values of under-provisioning, which can result in service degradation and potential dissatisfaction among end-users.

\begin{figure}[t!]
\centering
\includegraphics[width=1\columnwidth]{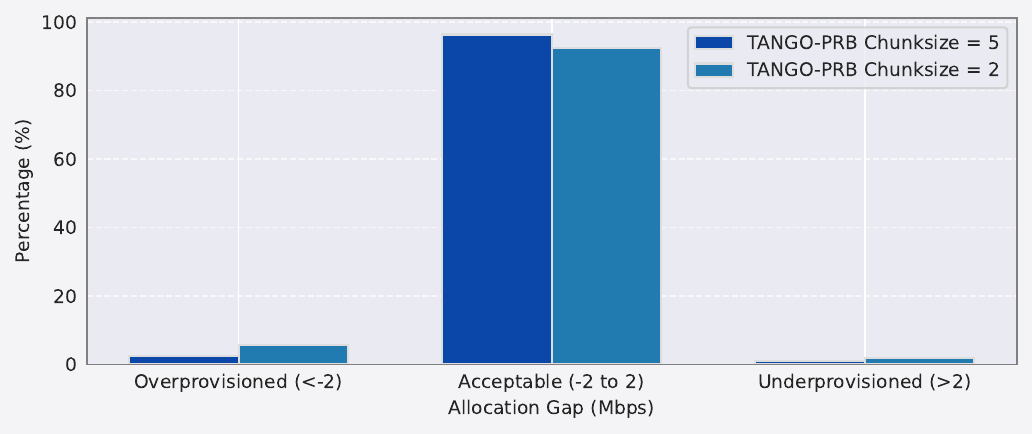}
\caption{Evaluation of resource allocation precision in the TANGO algorithm using different PRB chunk sizes (action space dimension) during the inference phase.}
\label{Sec:Scalability_Infernce}
\vspace{-.6cm}
\end{figure}

\subsection{Scalability Metric}
There are various approaches to consider when evaluating the scalability of a DRL algorithm~\cite{scaladrlLiu, scaladrlAndriotis, scaladrlShih}. One effective measure is to expand the size of the action space, which refers to the range or number of potential actions an agent can take. This approach increases complexity and requires the algorithm to handle a larger set of possible decisions. By doing so, the algorithm's ability to maintain or improve performance under increased complexity is put to the test. It also evaluates the efficiency of its learning and decision-making processes in more demanding scenarios. As such, the ability of a DRL algorithm to effectively manage larger action spaces is a critical indicator of its scalability and robustness. 

As Figure~\ref{Sec:Scalability_Infernce} demonstrates that TANGO algorithm can handle complex decision-making with action spaces with chunk size 2. Indeed, using a chunk size of 2 increases the action space dimension of TANGO compared with using a chunk size of 5. Here, the action space for chunk size 2 is 2.6$\times$ larger than the action space for chunk size 5. Despite the complexity, TANGO performs well, indicating its advanced ability to optimize decisions across a denser action space. This precision enables it to accurately make resource allocation adjustments, which is necessary when even minor discrepancies in resource distribution can significantly affect system performance. The TANGO algorithm's ability to maintain high performance even with dense and detailed decision-making (chunk size 2) demonstrates its scalability. This feature is precious in large-scale network environments that require high precision and adaptability. Moreover, the success in managing a more substantial number of decision points without a drop in performance highlights its adaptability, making it capable of responding appropriately to both minor and significant changes in demand. TANGO, with a broader range of possible actions, i.e., chunk size 5, benefits from a simplified learning process by dealing with a smaller set of actions. This can result in faster learning and convergence to effective strategies, especially when demand changes are not highly detailed, and a coarser adjustment can still efficiently meet the resource allocation needs.

\subsection{XAI Metric}
Table~\ref{table:nodes} highlights the feature importance and uncertainty values for different nodes in the TANGO model during the inference phase. The results provide insightful information on how the model perceives and uses the features associated with each node. Starting with Node 0, the table shows that both features are maximally important. This indicates that all the available features of Node 0 are crucial for the model's predictions. Regarding feature uncertainty, the values are relatively low (0.1605 and 0.1140), which signifies high confidence in their importance and stability in influencing the model's output. The results of Node 1 reveal a mixed importance, with one feature being maximally important and the other considerably less so. This suggests a disparity in how the features contribute to the model's decision-making process. The uncertainty values are moderately low but slightly higher than Node 0, especially for the less important feature. This implies less consistency in the influence of the second feature across different scenarios. For Node 2, the table indicates that one feature is minimally vital while the other is maximally important. Such a disparity indicates that the model relies heavily on specific characteristics of the node while almost ignoring others. The uncertainty is moderately high for both features, suggesting the model's reliance on the high-importance feature can be subject to conditions or context. Lastly, the table shows that Node 3 has a significant divergence in feature importance, with one being relatively essential and the other not at all. This indicates a very selective dependence on the features of Node 3. The uncertainty is high for the non-important feature, which contrasts with relatively low uncertainty for the important one. This reflects the model's variable confidence in how these features impact the outcomes.

Table~\ref{table:edge-attributes-comparison} provides valuable insights into the model's behavior and reliability during both the beginning of training and inference phases, which is crucial for XAI. The table shows how the model values certain connections, referred to as \emph{edges}, differently in each phase and how confident it is about those valuations. During the training phase, the edge $0 \rightarrow 1$ had maximum importance but no importance during inference, with a significant drop in uncertainty. This suggests that the model overfitted this connection during training but recognized it as irrelevant when presented with new data. On the other hand, the edge $0 \rightarrow 2$ had relatively low importance during training, but during inference, it became critical with maximum importance and consistently low uncertainty. This shift highlights an underutilized connection during training that became crucial when the model encountered new data, showing the model's adaptability. Overall, analyzing edges with significant changes in importance can help optimize model performance and improve its ability to handle new data.

\begin{table}[t!]
\centering
\caption{Node Attributes in Inference Phase}
\label{table:nodes}
\begin{tabular}{|c|c|c|}
\hline
\textbf{Node} & \textbf{Feature Importance} & \textbf{Feature Uncertainty} \\
\hline
Node 0 & [1.0, 1.0] & [0.1605, 0.1140] \\
Node 1 & [1.0, 0.4786] & [0.1656, 0.1478] \\
Node 2 & [0.2024, 1.0] & [0.1805, 0.1561] \\
Node 3 & [0.7958, 0.0] & [0.1271, 0.2168] \\
\hline
\end{tabular}
\end{table}

\begin{table}[ht]
\centering
\caption{Edge Attributes Comparison}
\label{table:edge-attributes-comparison}
\begin{tabular}{|c|cc|cc|}
\hline
\multirow{2}{*}{\textbf{Edge}} & \multicolumn{2}{c|}{\textbf{Training Phase}} & \multicolumn{2}{c|}{\textbf{Inference Phase}} \\ \cline{2-5} 
                               & \textbf{Importance} & \textbf{Uncertainty} & \textbf{Importance} & \textbf{Uncertainty} \\ 
\hline
0 $\rightarrow$ 1 & 1.0000 & 1.0000 & 0.0000 & 0.1196 \\
0 $\rightarrow$ 2 & 0.1885 & 0.6331 & 1.0000 & 0.1226 \\
0 $\rightarrow$ 3 & 0.3813 & 0.9956 & 0.9618 & 0.1506 \\
1 $\rightarrow$ 0 & 0.1634 & 0.8188 & 0.9452 & 0.2286 \\
1 $\rightarrow$ 2 & 0.0000 & 1.0000 & 0.0000 & 0.1546 \\
1 $\rightarrow$ 3 & 1.0000 & 0.9987 & 0.0000 & 0.1831 \\
2 $\rightarrow$ 0 & 0.2367 & 0.9254 & 0.0000 & 0.1994 \\
2 $\rightarrow$ 1 & 0.0000 & 1.0000 & 0.0000 & 0.1436 \\
2 $\rightarrow$ 3 & 0.0000 & 0.8972 & 0.8080 & 0.1993 \\
3 $\rightarrow$ 0 & 0.4581 & 1.0000 & 0.8290 & 0.1594 \\
3 $\rightarrow$ 1 & 0.0000 & 1.0000 & 0.0000 & 0.2163 \\
3 $\rightarrow$ 2 & 0.0323 & 0.9944 & 0.0000 & 0.1030 \\
\hline
\end{tabular}
\end{table}
During inference, Edge $0 \rightarrow 3$ and Edge $1 \rightarrow 0$ both exhibited an increase in importance and a decrease in uncertainty, indicating that these connections are essential for generalizing learned patterns to new data with increased confidence in their influence. On the other hand, inference showed an increasing importance in Edge $2 \rightarrow 3$ and Edge $3 \rightarrow 0$ with moderate uncertainty. These changes suggest that these edges capture relationships crucial for predictions under new conditions, possibly revealing hidden patterns not evident during training. Edges $1 \rightarrow 2$, $2 \rightarrow 1$, $3 \rightarrow 1$, and $3 \rightarrow 2$ consistently showed no or minimal importance in both phases, with varying levels of uncertainty. Therefore, these connections do not significantly influence the model's predictions, and the model is relatively confident about their lack of impact.

The analysis of edge attributes reveals how specific connections behave differently during both training and inference. For example, while Edge $0\rightarrow1$ holds the highest significance during training, its importance decreases significantly during inference, suggesting possible overfitting. Adapting and reconfiguring the GRL algorithm can help mitigate the excessive weighting of such connections during training. Overall, the analysis helps to make the model's decision-making process transparent by showing how it interprets and assesses different features and connections. We can create more robust models by adjusting the training methods to focus on the most influential and reliable features and connections. This can involve changing the training data and network states, refining the model's structure, or fine-tuning.

\section{Conclusion}
\label{sec:concusion_future}
This paper presents TANGO, an innovative GRL framework that enhances explainability and trustworthiness in AI-native 6G networks. TANGO integrates a Bayesian-GNN explainer and reasoner within a symbolic subsystem to provide in-hoc explanations during learning. This approach is crucial for mission-critical use-cases where traditional post-hoc XAI methods are insufficient. Our implementation of TANGO in a real-world testbed PoC for a gNB radio resource allocation problem demonstrated substantial improvements in network management operations. The framework effectively minimizes under- and over-provisioning of PRBs while penalizing decisions based on uncertain or unimportant edge-node relationships, underscoring its efficacy. Experimental results have shown that TANGO significantly expedites convergence compared to baseline and other benchmarks. Notably, TANGO achieved an impressive 96.39\% accuracy in optimal PRB allocation during the inference phase, surpassing the baseline by 1.22$\times$. Since TANGO shows higher complexity in some metrics compared to the benchmarks, we plan to address this in future work. Potential solutions include optimizing the algorithmic efficiency of TANGO's components, exploring more lightweight neural network architectures, or employing techniques like model pruning and quantization. This work paves the way for future research to explore the broader applicability of in-hoc explainability in various AI-driven 6G problems, enhancing trust and reliability in AI systems.


\vspace{-29pt}
\begin{IEEEbiography}
[{\includegraphics[width=1in,height=1.25in,clip,keepaspectratio]{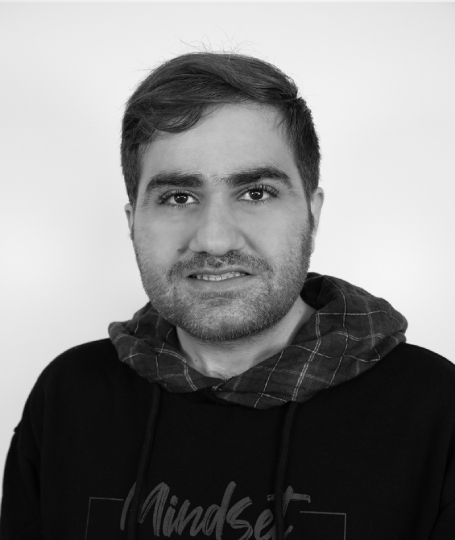}}]{Farhad Rezazadeh}~(Member, IEEE) received the Ph.D. degree (Excellent Cum Laude) in Signal Theory and Communications from Technical University of Catalonia (UPC), Barcelona, Spain. He is currently an R\&D engineer at the CTTC, Barcelona, Spain. He participated in 8 European and National 5G/B5G/6G R\&D projects with leading and technical tasks in the areas of Applied AI. 
He was a secondee at NEC Lab Europe and had scientific missions at TUM, Germany, TUHH, Germany and UdG, Spain. 
He is a Marie Sklodowska-Curie Ph.D. grantee and won 5 different IEEE/IEEE ComSoc grants, 2 European Cooperation in Science and Technology grants, and a Catalan Government Ph.D. Grant. He is an active member of ACM Professional, IEEE Young Professionals and IEEE Spain - Technical Activities and Standards with more than 29 top-tier journals/conferences and also book chapters. He actively serves as Organizing, Chair, Reviewer, and TPC member in IEEE and Guest Editor in Elsevier. He has over 130 verified reviews for peer-reviewed publications. He co-founded and coordinates the IEEE Trustworthy Internet of Things (TRUST-IoT) working group within the IEEE IoT Community.  
\end{IEEEbiography}

\vspace{-20pt}

\begin{IEEEbiography}[{\includegraphics[width=1in,height=1.25in,clip,keepaspectratio]{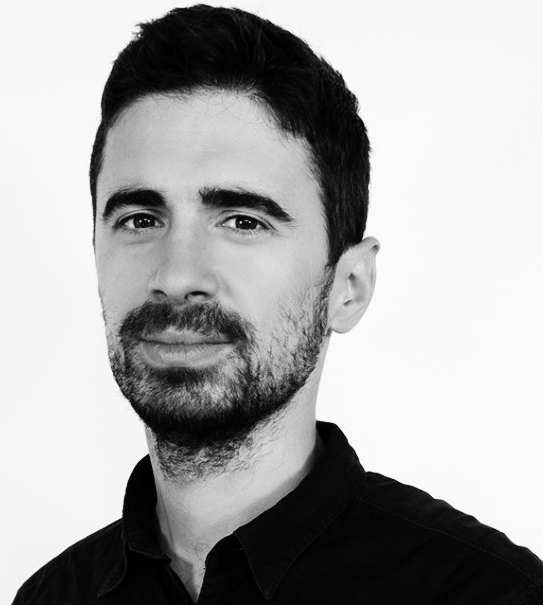}}]{Sergio Barrachina-Mu\~noz}~was born in Barcelona, Spain, in 1991. He received the B.S. degree in 2015, the M.S. degree in 2016, and the Ph.D. degree in 2021, all from Universitat Pompeu Fabra, Spain. He is currently a Senior Researcher and Technical Coordinator in the Services as NetworkS (SaS) research unit at the Centre Tecnologic de Telecomunicacions de Catalunya (CTTC). His research focuses on 5G/6G technologies, including next-generation networking, machine learning, orchestration, slicing, resource management, and experiment as a service/code.
\end{IEEEbiography}
\begin{IEEEbiography}[{\includegraphics[width=1in,height=1.25in,clip,keepaspectratio]{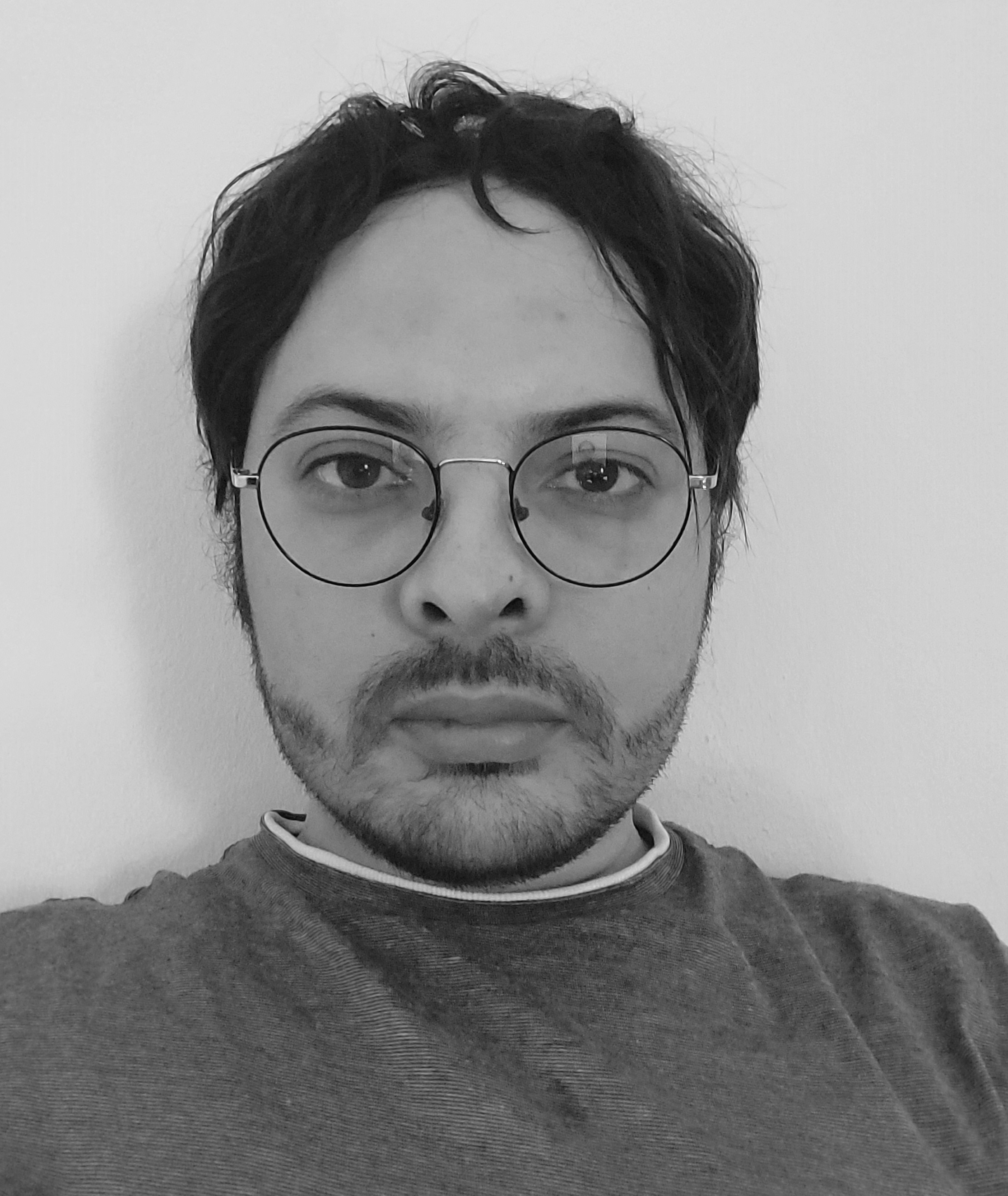}}]{Hatim Chergui}~(Senior Member, IEEE) is a Senior Researcher at i2CAT Foundation, Barcelona, Spain. He was the project manager of the H2020 MonB5G European project and a researcher at CTTC, Spain. He served as a RAN expert at both INWI and Huawei Technologies, Morocco. He has published more than 40 research papers in top-tier journals and conferences and has contributed to 1 European patent. He was the recipient of the IEEE ComSoc CSIM 2021 Best Journal Paper Award and the IEEE ICC 2020 Best Paper Award. He is an Associate Editor of IEEE Networking Letters and has been a Chair in several Workshops in IEEE Globecom and ICC. He has co-supervised PhD students in MSCA ITN projects 5GSTEPFWD and SEMANTIC.

\end{IEEEbiography}

\vspace{-20pt}
\begin{IEEEbiography}[{\includegraphics[width=1in,height=1.25in,clip,keepaspectratio]{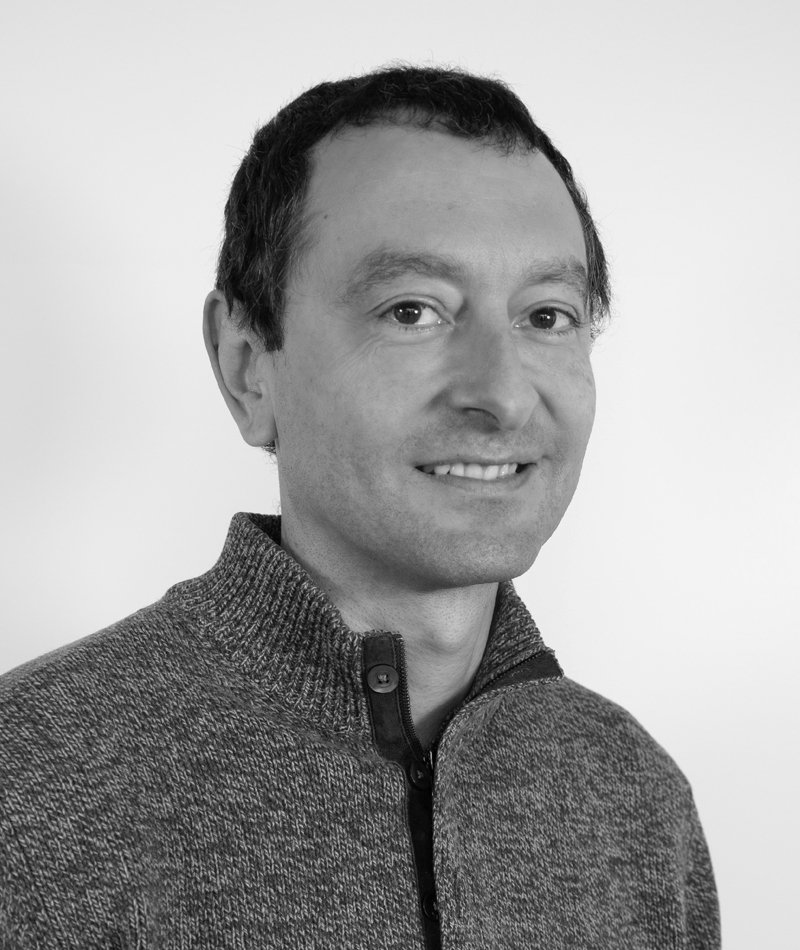}}]{Josep Mangues}~received the PhD degree in Telecommunications in 2003 from the Technical University of Catalonia (UPC). He is Research Director and Head of Services as networkS research unit of the CTTC. He has published 120+ journals/magazines and international conference papers through collaborations with multiple research groups. Since July 1997 he has participated in around 40 EU, Spanish, and industrial research projects in various roles (incl. PI). Interests include Cloud/Edge computing, NFV in mobile networks (incl. open RAN), data science/engineering, AI/ML-based service and network management automation and orchestration.
\end{IEEEbiography}

\vspace{-20pt}
\begin{IEEEbiography}[{\includegraphics[width=1in,height=1.25in,clip,keepaspectratio]{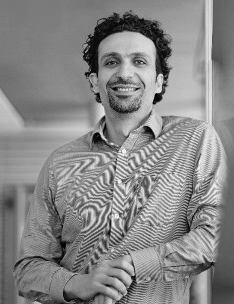}}]{Mehdi Bennis}~(Fellow, IEEE) is currently a tenured Full Professor with the Centre for Wireless Communications, University of Oulu, Finland,
and the Head of the Intelligent COnnectivity and Networks/Systems Group (ICON). He has published more than 200 research papers in international conferences, journals, and book chapters. His research interests include radio resource management, game theory, and distributed AI in 5G/6G
networks. He has been a recipient of several prestigious awards. He is an Editor of IEEE TRANSACTIONS ON COMMUNICATIONS and a specialty Chief Editor of Data Science for Communications and Frontiers in Communications and Networks journal.
\end{IEEEbiography}

\vspace{-20pt}
\begin{IEEEbiography}[{\includegraphics[width=1in,height=1.25in,clip,keepaspectratio]{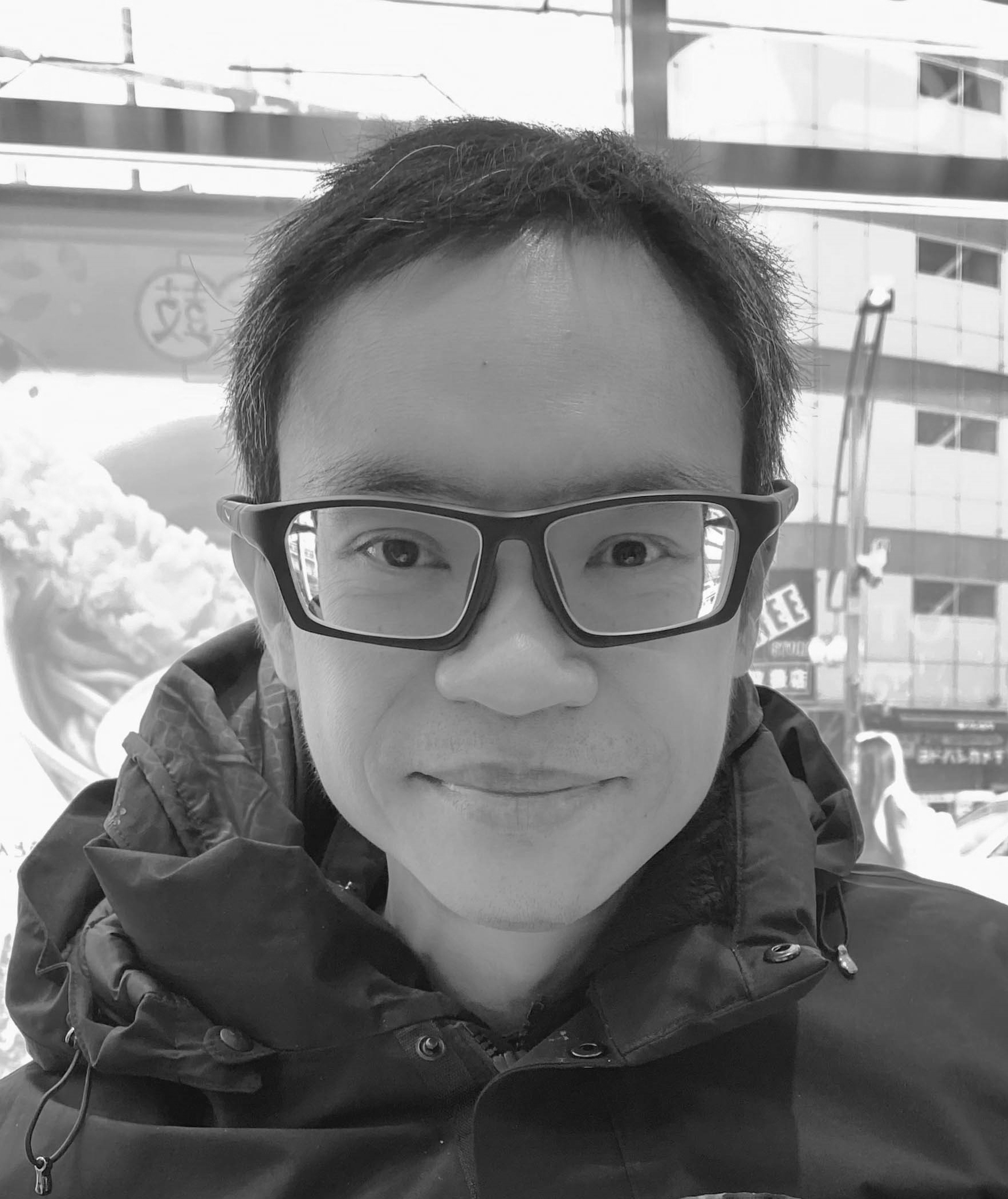}}]{Dusit Niyato}~Dusit Niyato (M'09-SM'15-F'17 IEEE) is a professor in the College of Computing and Data Science, at Nanyang Technological University, Singapore. He received B.Eng. from King Mongkuts Institute of Technology Ladkrabang (KMITL), Thailand and Ph.D. in Electrical and Computer Engineering from the University of Manitoba, Canada. His research interests are in the areas of mobile generative AI, edge intelligence, decentralized machine learning, and incentive mechanism design.
\end{IEEEbiography}
\vspace{-20pt}
\begin{IEEEbiography}[{\includegraphics[width=1in,height=1.25in,clip,keepaspectratio]{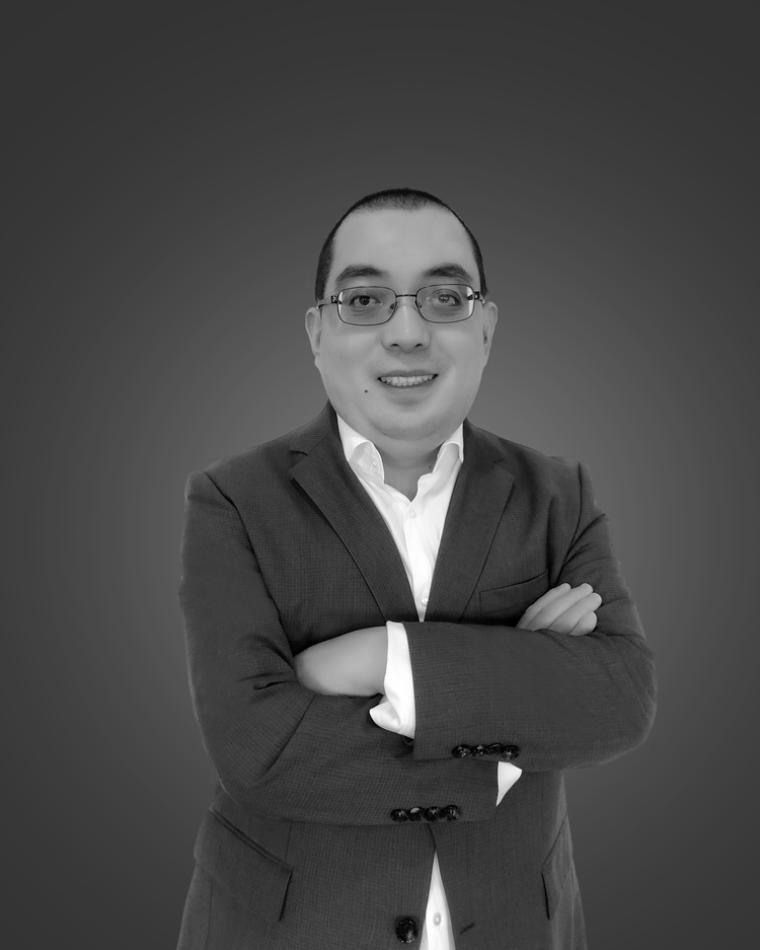}}]{Houbing Song}~(Fellow, IEEE) received the Ph.D. degree in electrical engineering from the University of Virginia, Charlottesville, VA, USA, in August 2012. He has been serving as an Associate Technical Editor, Associate Editor, and Guest Editor for IEEE journals and Transactions. He is an ACM Distinguished Member and an ACM Distinguished Speaker. He was a recipient of more than best paper awards from major international conferences, including the IEEE CPSCom-2019, the IEEE ICII 2019, the IEEE/AIAA ICNS 2019, the IEEE CBDCom 2020, the WASA 2020, the AIAA/IEEE DASC 2021, the IEEE GLOBECOM 2021, and the IEEE INFOCOM 2022.
\end{IEEEbiography}

\vspace{-20pt}
\begin{IEEEbiography}[{\includegraphics[width=1in,height=1.25in,clip,keepaspectratio]{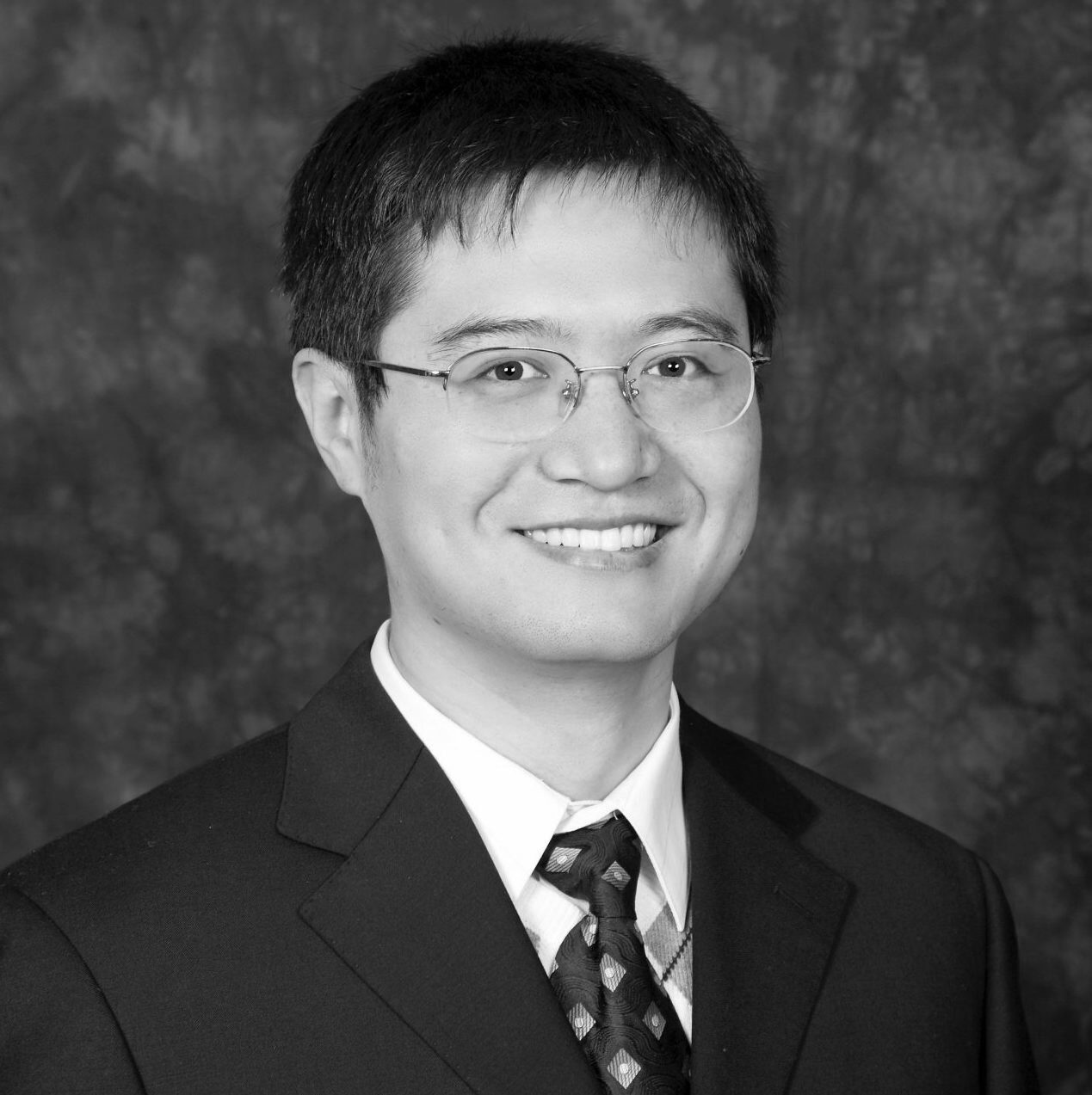}}]{Lingjia Liu}~(Senior Member, IEEE) received his Ph.D. degree in electrical
and computer engineering from Texas A\&M University, USA and his B.S. in electronic engineering
from Shanghai Jiao Tong University. Currently, he
is a professor in the Bradley Department of Electrical and Computer Engineering at Virginia Tech, USA. He is also serving as the Director of Wireless@Virginia Tech. His research interests include
machine learning for wireless communications, enabling technologies for 5G and beyond, mobile edge
computing, and Internet of Things.
\end{IEEEbiography}

\end{document}